\documentclass[3p]{elsarticle}
\pdfoutput=1
\usepackage{lineno, hyperref}
\modulolinenumbers[5]

\journal{Nuclear Physics B}

\begin{document}

\begin{frontmatter}

\title{Computing one-loop corrections to effective vertices with two scales in the EFT for Multi-Regge processes in QCD}

\author[mymainaddress,mysecondaryaddress]{M. A. Nefedov}
\ead{maxim.nefedov@desy.de}

\address[mymainaddress]{II. Institut f\"ur Theoretische Physik, Universit\"at Hamburg, Luruper Chaussee 149, 22761 Hamburg, Germany}
\address[mysecondaryaddress]{Samara National Research University, Moskovskoe Shosse 34, 443086 Samara, Russia}

\begin{abstract}
The computation of one-loop corrections to Reggeon-Particle-Particle effective vertices with two scales of virtuality is considered in the framework of gauge-invariant effective field theory for Multi-Regge processes in QCD. Rapidity divergences arising in loop integrals are regulated by ``tilted Wilson lines'' prescription. General analysis of rapidity divergences at one loop is given and necessary scalar integrals with one and two scales of virtuality are computed. Two examples of effective vertices at one loop are considered: the effective vertex of interaction of (space-like) virtual photon with one Reggeized and one Yang-Mills quark and the effective vertex of Reggeized gluon to Yang-Mills gluon transition with an insertion of the operator ${\rm tr}\left[G_{\mu\nu}G^{\mu\nu} \right]$ carrying the (space-like) off-shell momentum. All terms $\sim r^{\pm\epsilon}$ in the rapidity-regulator variable $r$ cancel between diagrams and only $\log r$-divergence is left. It is checked on several examples, that obtained results indeed allow one to reproduce Regge limit of one-loop QCD scattering amplitudes.
\end{abstract}


\end{frontmatter}


\section*{Introduction}

  The physics of Multi-Regge limit of scattering amplitudes in Yang-Mills theory (see e.g. Refs.~\cite{RevLipatov97, RevDelDuca95} for the review) continues to attract a lot of attention both form theoretical and phenomenological points of view. From the theory side, the list of applications of Multi-Regge limit includes studies~\cite{Bartels:2008ce, Bartels:2008sc, Bartels:2014jya} of corrections to the Bern-Dixon-Smirnov all-order ansatz~\cite{Bern:2005iz} for scattering amplitudes in ${\cal N}=4$-supersymmetric Yang-Mills theory and recently discovered connection~\cite{IRfactorRegge, Caron-Huot:2017zfo} between Multi-Regge asymptotics of scattering amplitudes and all-order structure of infrared divergences in {\it non-supersymmetric} QCD, summarized by dipole formula~\cite{Becher:2009cu, Gardi:2009qi, Becher:2009qa, Gardi:2009zv}.  

  From more phenomenological perspective, scattering amplitudes in Multi-Regge Kinematics (MRK) form a basis for the celebrated Balitsky-Fadin-Kuraev-Lipatov (BFKL)~\cite{BFKL1, BFKL2, BFKL3} evolution equation, which allows one to resum the higher-order corrections to scattering amplitudes or observable cross-sections in QCD, enhanced by powers of $\log s/(-t)$, where $s$ is a (partonic) squared center-of-mass energy and $t$ is the typical $t$-channel momentum transfer in the process under consideration. In QCD, the kernel of BFKL equation is computed up to the Next-to-Leading order(NLO) level~\cite{NLO-BFKL,NLOCiafaloni1,NLOCiafaloni2}, but it's direct phenomenological application is impossible because NLO correction turns out to be numerically large. Logarithms coming from the Dokshitzer-Gribov-Lipatov-Altarelli-Parisi~\cite{DGLAP1, DGLAP2, DGLAP3} evolution in transverse-momentum scale~\cite{Salam98} and running of QCD coupling~\cite{Brodsky:1998kn} has been identified as the main sources of perturbative instability of the BFKL kernel and their consistent resummation is one of the central problems of BFKL phenomenology today, see e.g. Refs.~\cite{RGIBFKL, ChVeraRGI1, Ball:2017otu, Marzani:2007gk} for the recent work in this direction.  

  The property of exponentiation of loop corrections enhanced by $\log s/(-t)$ is known under the name of {\it Reggeization} of quarks and gluons in QCD and it is proven up to Next-to-Leading Logarithmic approximation (NLLA)~\cite{ReggeProofLLA, ReggeProofNLLA, BogFad06}. It is natural to ask for the systematic tool, which makes this property of QCD scattering amplitudes manifest. The gauge-invariant Effective Field Theory (EFT) for Multi-Regge processes in QCD~\cite{Lipatov95, LV} is such a tool. In the original papers~\cite{Lipatov95, LV}, the Reggeization of gluon and quark was shown to hold in a framework of EFT in the LLA. Computing loop corrections in this EFT one finds a new type of divergences of loop integrals, the so-called {\it rapidity divergences}, which will be discussed in the forthcoming sections of the present paper. Similar divergences also arise in the context of Soft-Collinear Effective Theory (SCET)\footnote{On the possibility to include BFKL effects and quark Reggeization into framework of SCET see e.g. Refs.~\cite{Rothstein:2016bsq, Moult:2017xpp}.}~\cite{Becher:2011dz, Chiu:2012ir} and Transverse-Momentum-Dependent (TMD) factorization~\cite{CollinsQCD, Collins:2011ca} and several prescriptions to regularize them, such as analytic regularization~\cite{Becher:2011dz}, $\delta$-regularization~\cite{Echevarria:2015byo} and ``tilted Wilson lines'' regularization~\cite{CollinsQCD, Collins:2011ca} where introduced in the literature.  Among mentioned prescriptions, only the last one does not modify the standard definition of the Wilson lines, and therefore it can be relatively straightforwardly applied to the High-Energy EFT~\cite{Lipatov95, LV}. First such applications where performed in the Refs.~\cite{Hentschinski:2011tz, Chachamis:2012cc, Chachamis:2012gh} where one-loop corrections to the propagator of Reggeized gluon and effective vertices of interaction of on-shell quark and gluon with one Reggeized gluon where computed in the framework of EFT, and found to be consistent with the results obtained earlier from QCD. Later, the two-loop Regge trajectory of a gluon was extracted from a rapidity-divergent part of the two-loop correction to a Reggeized gluon propagator in the EFT~\cite{Chachamis:2013hma} and it coincides with known QCD result. Also, the calculation of the one-loop correction to the propagator of Reggeized quark ($Q$) and vertex of interaction of on-shell photon with one Reggeized and one QCD quark ($Q\gamma q$-vertex) has been done in the Ref.~\cite{gaQq-real-photon} and it was shown, that this results allow one to reproduce the Multi-Regge asymptotics of the positive-signature part of the one-loop QCD amplitude for $\gamma\gamma\to q\bar{q}$ process, thus checking the consistency of High-Energy EFT in the Reggeized quark sector. Further development of computational techniques within this approach, in particular going beyond LLA or considering quantities with more than one scale of virtuality, is required for the EFT-approach to be instrumental for the solution of above-mentioned problems of BFKL-physics.

  In the present paper we deal with the problem of computation of one-loop corrections to Reggeon-Particle-Particle vertices, containing one additional scale of virtuality, besides the transverse momentum of the Reggeized parton. We will consider two examples of such quantities at one loop: the $Q\gamma^\star q$-vertex with a space-like off-shell photon ($\gamma^\star$) and effective vertex of Reggeized gluon ($R$) to Yang-Mills gluon ($g$) transition with an insertion of gauge-invariant local operator ${\cal O}(x)=-{\rm tr}\left[G_{\mu\nu}(x) G^{\mu\nu}(x) \right]/2$ carrying the space-like off-shell four-momentum, which we call the $R{\cal O}g$-vertex. Besides methodological significance of this calculations, this results will also become an integral part of virtual NLO corrections to the structure functions of Deep Inelastic Scattering (DIS) at NLO of Parton Reggeization Approach(PRA) to the multi-scale hard processes at Hadron Colliders, see Ref.~\cite{NS_PRA} for the introduction. The real NLO correction for DIS hard subprocess with Reggeized gluon in the initial state has been computed in PRA in Refs.~\cite{NS_DIS1, NS_DIS2}, where the problem of definition of unintegrated Parton Distribution Function at NLO is also discussed.

  The present paper is organized as follows. Introduction to the formalism of High-Energy EFT and problem of rapidity divergences is given in the Sec.~\ref{Sec:EFT}. In Sec.~\ref{Sec:RDs-gen} we perform the general analysis of rapidity divergences in the one-loop integrals with one and two external Reggeon lines and in Sec.~\ref{Sec:Ints} we compute necessary scalar integrals. The calculation of one-loop corrections to $Q\gamma^\star q$ and $R{\cal O}g$-vertices proceeds in the Sec.~\ref{Sec:1-loop-vert}, where the fate of spurious power-divergent terms is traced and it is shown, that they cancel, leaving only sigle-logarithmic rapidity divergence, as it should be according to Reggeization hypothesis. In Sec.~\ref{Sec:comp-QCD} and \ref{Sec:Imag} we compare results obtained in the EFT with direct QCD calculations. With the use of obtained one-loop corrections to effective vertices we reconstruct the Multi-Regge asymptotics of the one-loop amplitude of DIS on the on-shell photon target and the DIS-like process on the on-shell gluon target, driven by the operator ${\cal O}(x)$. The one-Reggeon exchange contributions are considered in Sec.~\ref{Sec:comp-QCD} and they contribute both to real and imaginary parts of amplitude at this order. In the Sec.~\ref{Sec:Imag} two-Reggeon contributions in the EFT are discussed, which contribute only to the imaginary part at one loop. Finally, our conclusions are summarized in the Sec.~\ref{Sec:Conclusions} and in the \hyperlink{Sec:Appendix:A}{Appendix}, which we refer to in Sec.~\ref{Sec:EFT}, the comparison of pole prescriptions for the $O(g_s^2)$ and $O(g_s^3)$ induced vertices arising from the Hermitian version of the Reggeon-gluon interaction~\cite{RevLipatov97, BondZubkov} with results of Ref.~\cite{MH_PolePrescr} is presented.      

\section{Gauge-invariant Effective Field Theory for Multi-Regge processes in QCD}
\label{Sec:EFT}

\begin{figure}
\begin{center}
\includegraphics[width=0.6\textwidth]{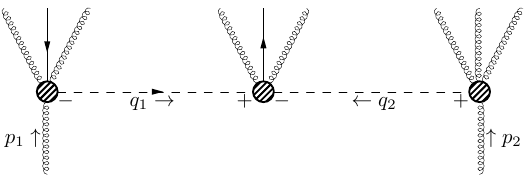}
\end{center}
\caption{An example of QMRK process with three clusters of final-state partons highly separated in rapidity. The Reggeized quark and Reggeized gluon in the $t$-channels are denoted by dashed lines. The effective vertices are denoted by by shaded circles. }\label{Fig:QMRK}
\end{figure}

  In the limit of MRK for the $2\to 2+n$ partonic subprocess, all $2+n$ final-state particles are highly separated in rapidity, while their transverse momenta are small in comparison with the collision energy. The kinematic situation when one can identify a few groups of final-state partons each of which occupy a relatively small rapidity interval in comparison with rapidity gaps between groups is called Quasi-Multi-Regge kinematics(QMRK). If one introduces the invariant mass $s_{ij}\sim e^{\Delta y_{ij}}$ for each pair of highly-separated final-state partons $i$ and $j$ in MRK (or each pair of clusters of partons in QMRK), and (anti-)symmetrizes the full QCD scattering amplitude w.r.t. substitutions $s_{ij}\leftrightarrow u_{ij}\simeq -s_{ij}$, obtaining in such a way the amplitude with {\it definite signature} in each of the $s_{ij}$-channels, then in the QMRK limit, such an object is shown~\cite{ReggeProofLLA, ReggeProofNLLA, BogFad06} to factorize into a product of certain universal gauge-invariant parts. An example of such factorization can be given with the help of QMRK process, depicted in the Fig.~\ref{Fig:QMRK}. Here the QMRK asymptotics of the part of total QCD amplitude with positive signature in $s_{12}=(p_1+q_2)^2$-channel and negative signature in $s_{23}=(p_2+q_1)^2$-channel can be represented as a product of three {\it effective vertices}, corresponding to the production of three clusters of final-state partons, and $t$-channel propagators of {\it Reggeized gluons} ($R_{\pm}$) and {\it Reggeized quarks} ($Q_{\pm}$), collectively named as {\it Reggeons}. The Reggeized gluon is a scalar particle in the adjoint representation of the color group $SU(N_c)$, while the Reggeized quark is the Dirac fermion in the fundamental representation of the color group. The propagators of Reggeons are dressed by Regge factors $\left( (s_{12}/s_0)^{\omega_q(q_1^2)}+(-s_{12}/s_0)^{\omega_q(q_1^2)} \right)/2$ and $\left( (s_{23}/s_0)^{\omega_g(q_2^2)}+(-s_{23}/s_0)^{\omega_g(q_2^2)} \right)/2$, which determine the dependence of amplitude on energies $s_{12}$ and $s_{23}$, while effective vertices depend only on an arbitrary {\it rapidity-factorization scale} $s_0$, and $\omega_{g/q}(q_i^2)$ are functions of $q_i^2$ and $\alpha_s$ called {\it gluon}({\it/quark}) {\it Regge trajectories.} These functions contain explicit infrared divergences and in QCD they are currently known up to $O(\alpha_s^2)$.    

 It is most convenient to describe the QMRK if one introduces Sudakov decomposition for an arbitrary four-vector:
\begin{equation}
k^\mu=\frac{1}{2}\left(k_+n_-^\mu + k_-n_+^\mu \right) + k_T^\mu,  \label{Eq:Sud-dec}
\end{equation}
where\footnote{The reader should notice, that unlike to the notation of e.g. Ref.~\cite{Lipatov95}, we do not distinguish between covariant and contravariant light-cone components, so that the position of indices $\pm$ in our formulas does not have any meaning. } $n^\mu_-=(n^-)^\mu=p_1^\mu/\sqrt{s}$, $n^\mu_+=(n^+)^\mu=p_2^\mu/\sqrt{s}$, $s=(p_1+p_2)^2$, $p_1^2=p_2^2=0$, so that $n_{\pm}^2=0$, $n_+n_-=2$ and $k_{\pm}=k^{\pm}=n_{\pm}k$, $n_\pm k_T=0$. In the center-of-mass frame of $p_1$ and $p_2$, the relation $k^{\pm}=k^0\pm k^3$ holds, and for the square of four-vector one has $k^2=k_+k_- - {\bf k}_T^2$ while rapidity of a particle is defined as $y=\log (k_+/k_-)/2$. By definition, momentum $p_1$ has large $p_1^+$ component, $p_2$ has large $p_-$ component and in QMRK situation depicted in Fig.~\ref{Fig:QMRK}, only small fractions of this large components are transferred to the cluster at central rapidities:
\[
z_1=\frac{q_1^+}{p_1^+}\ll 1,\ \ z_2=\frac{q_2^-}{p_2^-}\ll 1.
\]
  Consequently, on entrance to the central production vertex, the following scaling relations for components of momenta $q_{1,2}$ holds:
 \begin{eqnarray}
  |{\bf q}_{T1}|\sim q_1^+\sim O(z_1)\ll q_1^-\sim O(z_1^2), \nonumber \\
  |{\bf q}_{T2}|\sim q_2^-\sim O(z_2)\ll q_2^+\sim O(z_2^2), \label{Eq:MRK-scalings}
  \end{eqnarray}
so that one (so-called ``large'') light-cone component of these momenta is of the same order as corresponding transverse momentum, while the other one is negligible and momenta $q_{1,2}$ are necessarily off-shell: $q_{1,2}^2\simeq -{\bf q}_{T1,2}^2$.

 Effective vertices describe interaction of Reggeons and ordinary QCD partons. They are gauge-invariant independently from each-other, despite the fact that Reggeons connected to them are off-shell. This is a nontrivial property because in general, off-shell Green's functions in QCD are not gauge invariant. Gauge-invariance dictates the specific form of interaction of particles and Reggeons in the EFT~\cite{Lipatov95, LV}, containing Wilson lines. 

 In QCD {\it at tree level,} the above-described {\it Regge-pole factorization} is true up to corrections, suppressed by powers of $s_{ij}$. If loop corrections are taken into account, it holds in the LLA (when one takes all the terms proportional to $\alpha_s^n \log^n s_{ij}$ from all orders of perturbation theory) and in the NLLA (when one adds also $\sim\alpha_s^{n+1} \log^n s_{ij}$ terms). Beyond NLLA, Regge-pole factorization is violated by Regge cut contribution~\cite{DelDuca:2001gu, Caron-Huot:2017fxr, Fadin:2017nka} and nonlinear effects of interaction of multiple Reggeons in the $t$-channel, but it should be possible to formulate both of this effects in the language of High-Energy EFT, see e.g. recent Ref.~\cite{Hentschinski:2018rrf} on the non-linear effects.  

 The construction of High-Energy EFT~\cite{Lipatov95, LV} proceeds as follows. The axis of rapidity is sliced into a few intervals, corresponding to clusters of particles, highly separated in rapidity. For the QCD gluon and quark fields, ``living'' in each interval, the separate copy of QCD Lagrangian:
\begin{equation}
 L_{\rm QCD}(A_\mu, \psi_q)=-\frac{1}{2}{\rm tr}\left[G_{\mu\nu}G^{\mu\nu}\right]+\sum\limits_{q=1}^{n_F}\bar{\psi}_q (i\hat{D}-m_q)\psi_q, \label{Eq:L-QCD}
\end{equation} 
is defined, where $\hat{D}=\gamma_\mu D^\mu$, $D_\mu=\partial_\mu+ig_s A_\mu$ is the covariant derivative, $A_\mu=A_\mu^a T^a$, $T^a$ are the (Hermitian) generators of the fundamental representation of $SU(N_c)$, $g_s$ is the coupling constant of QCD and $G_{\mu\nu}=-i\left[D_\mu,D_\nu\right]/g_s$ is the Yang-Mills field-strength tensor.  The complete Lagrangian of the EFT takes the form:
   \begin{eqnarray}
 & L_{\rm eff}= L_{\rm kin.} + \sum\limits_i \left[ L_{\rm QCD}(A_\mu^{[y_i, y_{i+1}]}, \psi_q^{[y_i, y_{i+1}]})\right. \nonumber \\
 & \left. + L_{\rm Rg}(A_\mu^{[y_i, y_{i+1}]}, R_+, R_-)  + L_{\rm Qqg}(A_\mu^{[y_i, y_{i+1}]}, \psi_q^{[y_i, y_{i+1}]}, Q_+, Q_-)  \right],   \label{Eq:LEFT}
  \end{eqnarray}
  where the index $[y_i, y_{i+1}]$ of the field means, that the real part of rapidity of it's momentum modes is restricted to lie within the interval $y_i\leq {\rm Re}(y) \leq y_{i+1}$. Kinetic part of the Lagrangian takes the form:
   \begin{equation}
  L_{\rm kin.}=4{\rm tr}\left[ R_+ \partial_T^2 R_- \right] + \overline{Q}_- \left(i\hat{\partial}_T\right) Q_+ + \overline{Q}_+ \left(i\hat{\partial}_T\right) Q_-, \label{Eq:L_kin}
  \end{equation}
  which leads to non-diagonal bare propagators, connecting $R_+$-field with $R_-$: $-i/(2{q}_T^2)$ and $Q_+$-field with $Q_-$: $i\hat{q}_T/{q_T^2}$, where $q_T$ is the transverse part of the four-momentum of the Reggeon. Due to MRK-conditions (\ref{Eq:MRK-scalings}) fields of Reggeized gluons $R_{\pm}$ and quarks $Q_{\pm}$ are subject to the following kinematic constraints:
  \begin{eqnarray}
  \partial_+ R_- = \partial_- R_+ = 0, \label{Eq:kin-constr-R}\\
  \partial_+ Q_- = \partial_- Q_+ = 0, \label{Eq:kin-constr-Q}
  \end{eqnarray}
  where $\partial_\pm=n_{\pm}^\mu\partial_\mu=2\partial/\partial x_{\mp}$. Also, to remove the Dirac structures, associated with ``small'' light-cone component of Reggeon momentum, the following constraints are applied to the spinor structure of the fields of Reggeized quarks~\cite{LV}:
\begin{equation}
  \hat{n}_\pm Q_{\mp} =0,\ \overline{Q}_{\pm} \hat{n}_{\mp}=0. \label{Eq:kin_cons_Q2}
\end{equation}
On the level of Feynman rules the latter constraints are implemented by adding the following projectors to propagators of Reggeized quarks:
\begin{equation}
\hat{P}_{\pm}=\frac{1}{4}\hat{n}_{\mp} \hat{n}_{\pm}. \label{Eq:Proj-Qpm}
\end{equation}

  Kinematic constraints (\ref{Eq:kin-constr-R}) and (\ref{Eq:kin-constr-Q}) are not invariant w.r.t. local gauge transformations, and therefore fields $R_{\pm}$ and $Q_{\pm}$ have to be gauge-invariant by themselves. This requirement mostly fixes the form of gauge-invariant interaction between Reggeons and QCD partons. The Hermitian form~\cite{RevLipatov97,BondZubkov} of the Lagrangian of interaction of Reggeized and QCD gluons~\cite{Lipatov95} can be written as follows:
   \begin{equation}
  L_{Rg}(x)=\frac{i}{g_s}{\rm tr}\left[R_+(x) \partial_\rho^2 \partial_- \left(W_x[A_-]-W_x^\dagger[A_-]\right) + R_-(x) \partial_\rho^2\partial_+ \left(W_x[A_+]-W_x^\dagger[A_+]\right) \right], \label{Eq:L-Rg}
  \end{equation}
  where $W_x[A_{\pm}]$ is (past-pointing) half-infinite Wilson line, stretching in the $(+)$ or $(-)$ light-cone direction from the point $x$:
\begin{eqnarray}
  W_x[A_{\pm}]&=& P\exp\left[\frac{-ig_s}{2} \int\limits_{-\infty}^{x_{\mp}} dx'_{\mp} A_{\pm}\left(x_{\pm}, x'_{\mp}, {\bf x}_{T}\right)  \right] \nonumber \\
  &=& 1-ig_s\left(\partial_\pm^{-1}A_{\pm} \right) + (-ig_s)^2\left(\partial_\pm^{-1}A_{\pm}\partial_\pm^{-1}A_{\pm}\right)+\ldots , \label{Eq:WL-def} \\
  \overline{W}_x[A_{\pm}]&=& P\exp\left[\frac{-ig_s}{2} \int\limits^{+\infty}_{x_{\mp}} dx'_{\mp} A_{\pm}\left(x_{\pm}, x'_{\mp}, {\bf x}_{T}\right)  \right] \nonumber \\
  &=& 1-ig_s\left(\bar{\partial}_\pm^{-1}A_{\pm} \right) + (-ig_s)^2\left(\bar{\partial}_\pm^{-1}A_{\pm}\bar{\partial}_\pm^{-1}A_{\pm}\right)+\ldots , \label{Eq:WLbar-def}
  \end{eqnarray}
where we have defined also the future-pointing Wilson line $\overline{W}_x[A_{\pm}]$ and operators $\partial^{-1}_{\pm}$ or $\bar{\partial}^{-1}_{\pm}$ act as $\partial^{-1}_{\pm}f(x)=\int\limits_{-\infty}^{x^{\mp}} dx'_{\mp}/2\ f(x_{\pm},x'_{\mp},{\bf x}_T)$ and $\bar{\partial}^{-1}_{\pm}f(x)=\int\limits^{+\infty}_{x^{\mp}} dx'_{\mp}/2\ f(x_{\pm},x'_{\mp},{\bf x}_T)$ so that on the level of Feynman rules, they correspond to Eikonal factors with definite $i\varepsilon$-prescription:$-i/(k_{\pm}+i\varepsilon)$ and $i/(k_{\pm}-i\varepsilon)$ respectively.  
  
  Lagrangian (\ref{Eq:L-Rg}) generates an infinite sequence of {\it induced vertices} of interaction of Reggeized gluon with $n$ QCD gluons. The simplest of them is $R_+g$-transition vertex, corresponding to $O(g_s^0)$-term in (\ref{Eq:L-Rg}):
\begin{eqnarray}
L_{Rg}\supset \frac{i}{g_s}{\rm tr}\left[ R_+ \partial_\rho^2 \partial_- (-2i g_s) \partial_-^{-1} A_- \right] \rightarrow \Delta_{+\mu}^{ab}(q)=(-iq^2)n^-_\mu\delta_{ab}, 
\end{eqnarray} 
where $q$ is (incoming) momentum of the Reggeon, factor $2$ in circular brackets corresponds to taking into account both $W$ and $W^\dagger$ terms in (\ref{Eq:L-Rg}) and $R_-g$ vertex $\Delta_{-\mu}^{ab}$ can be obtained by the $-\leftrightarrow +$ replacement in the obtained expression.

 The $R_+gg$ induced vertex is generated by $O(g_s)$-term in (\ref{Eq:L-Rg}):
\begin{eqnarray}
L_{Rg}\supset \frac{i}{g_s}{\rm tr}\left[ R_+ \partial_\rho^2 \partial_- (-g_s^2)\left( T^{b_1}T^{b_2}-T^{b_2} T^{b_1} \right) \partial_-^{-1}A_-^{b_1}\partial_-^{-1}A_-^{b_2} \right] = -i g_s  \frac{if^{ab_1b_2}}{2} R^a_+ \partial_\rho^2 A_-^{b_1} \partial_-^{-1} A_-^{b_2} \nonumber \\
\rightarrow \Delta_{+\mu_1 \mu_2}^{ab_1 b_2}(q,k_1) = g_s (n^-_{\mu_1} n^-_{\mu_2}) \frac{q^2}{2} \left(\frac{f^{a b_1 b_2}}{k_2^-+i\varepsilon} + \frac{f^{a b_2 b_1}}{k_1^-+i\varepsilon}  \right) = g_s q^2 (n^-_{\mu_1} n^-_{\mu_2}) \frac{f^{ab_1b_2}}{2} \left( \frac{1}{k^-_1+i\varepsilon} + \frac{1}{k^-_1-i\varepsilon} \right),  \label{eq:Rgg-ind-vert}
\end{eqnarray}
where $k_1$ and $k_2$ are (incoming) momenta of the gluons, the term $T^{b_2}T^{b_1}$ in the first line came from the $W^\dagger$ and in the second line we have used the condition $k_1^-+k_2^-=0$ which follows from the constraint (\ref{Eq:kin-constr-R}). One can see, that the PV-prescription for the $1/k^-$-pole in the $Rgg$ induced vertex, which was advocated in Ref.~\cite{MH_PolePrescr}, follows form the Lagrangian of $Rg$-interaction in the form~(\ref{Eq:L-Rg}).  Pole prescription for higher-order induced vertices is more complicated and constrains also the color structure of the vertex. In the Ref.~\cite{MH_PolePrescr} the $i\varepsilon$ structure of induced vertices was derived from the requirement for the exchanges of Reggeized gluons in $t$-channels to have a negative signature, and this prescription was later used in the calculations of Refs.~\cite{Hentschinski:2011tz, Chachamis:2012cc, Chachamis:2012gh, Chachamis:2013hma}. It turns out, that induced vertices which satisfy all conditions introduced in Ref.~\cite{MH_PolePrescr} can be obtained directly from the Lagrangian~(\ref{Eq:L-Rg}), without any additional adjustments. At least, this statement had been verified by us in the orders $O(g_s^2)$ and $O(g_s^3)$, see the \hyperlink{Sec:Appendix:A}{Appendix}.

 The Lagrangian of interaction of Reggeized quarks with QCD quarks and gluons has the form~\cite{LV}:
  \begin{equation}
  L_{\rm Qqg}(x)= - \overline{Q}_-(x) \frac{i\hat{\partial}}{2} \left( W_x^\dagger [A_+]+\overline{W}_x[A_+]\right) \psi(x)  - \overline{Q}_+(x) \frac{i\hat{\partial}}{2} \left( W_x^\dagger [A_-] + \overline{W}_x[A_-] \right)\psi(x)  + {\rm h. c.}, \label{Eq:L-Qq}
  \end{equation}
  where ${\rm h.c.}$ denotes the Hermitian-conjugate terms. Working with Reggeized quarks it is most convenient to remove the $Q_{\pm}q$-transition vertex by the following shift of quark fields:
  \[
  \psi\to \psi + \hat{P}_+Q_+ + \hat{P}_-Q_-,\ \ \overline{\psi}\to \overline{\psi} + \overline{Q}_+ \hat{P}_- + \overline{Q}_- \hat{P}_+.
  \]
  After the shift, only vertices of the type $Q_{\pm}gq$, $Q_{\pm}ggq$, ..., $Q_+ g Q_-$, $Q_+ggQ_-$ e.t.c. are left in the Lagrangian. The $O(g_s)$ $Q_+gq$-vertex~\cite{FadinSherman76, FadinSherman77, BogFad06} is obtained from the following term in the sum of (\ref{Eq:L-Qq}) and (\ref{Eq:L-QCD}) after the shift:
\begin{eqnarray}
L_{\rm Qqg}+L_{\rm QCD}\supset i\overline{Q}_-\left[ ig_s \hat{A}+\frac{\hat{\overleftarrow{\partial}}}{2} \left(ig_s \partial^{-1}_- A_- - ig_s \bar{\partial}^{-1}_- A_-\right) \right] \psi \nonumber \\
 \rightarrow  g_sT^a\cdot \Gamma^{(0),a}_{-\mu}(q, k)=g_sT^a\left[ \gamma_\mu+ \frac{\hat{q} n_\mu^-}{2}\left( \frac{1}{k^-+i\varepsilon} + \frac{1}{k^- - i\varepsilon} \right)   \right], \label{Eq:Qgq-vert}
\end{eqnarray}
where $k$ is the (incoming) gluon momentum and index $(0)$ denotes, that this is the LO result in $g_s$, without any loop corrections. One can see, that PV-prescription for the simplest $Qgq$ effective vertex is obtained from the Lagrangian (\ref{Eq:L-Qq}). In the following expressions, we will denote the PV-prescription for the pole by square brackets:
\[
\frac{1}{[k^\pm]}=\frac{1}{2}\left( \frac{1}{k^-+i\varepsilon} + \frac{1}{k^- - i\varepsilon} \right).
\] 

  Analogously, one can obtain the Fadin-Sherman~\cite{FadinSherman76, FadinSherman77} $Q_+gQ_-$-vertex:
\begin{equation}
g_sT^a\cdot\Gamma^{(0),a}_{+\mu-}(q_1,k,q_2)=g_sT^a\left[ \gamma_\mu+ \frac{\hat{q}_1 n_\mu^-}{[k^-]} + \frac{\hat{q}_2 n_\mu^+}{[k^+]}  \right],\label{Eq:FS-vertex}
\end{equation}
where $q_1$ and $q_2$ are incoming momenta of $Q_+$ and $Q_-$, while $k$ is incoming gluon momentum.

  Below, we will also need the couplings of photons to Reggeized quarks. They can be obtained from vertices of interaction of Reggeized quarks with QCD quarks and gluons by replacement of corresponding factors $g_s T^a_{ij}$ by $ee_q\delta_{ij}$ where $e$ is the coupling constant of QED and $e_q$ is the quark charge in units of electron charge. 

  As it will be discussed in details in Sec.~\ref{Sec:RDs-gen}, in certain kinematics, loop integrals containing Eikonal propagators exhibit a new type of divergences, which are not regulated by the usual dimensional regularization. These are so-called {\it rapidity divergences}, and in the standard formulation of effective action (\ref{Eq:LEFT}) such divergences are regulated by the cutoffs on the real part of rapidity of loop momenta or momenta of real particles produced within the given cluster: $y_i\leq {\rm Re}(y) \leq y_{i+1}$ and they manifest themselves as terms proportional to $y_{i+1}-y_i$. Dependence on the regulators $y_i$ cancels between the contributions of neighboring clusters in each order of perturbation theory, building up the terms which can be finally resummed into Regge exponents $\exp[Y \omega_{q/g}(t)]$. 

  However, for practical calculations, the cutoff regularization is rather inconvenient, especially in the multi-scale case. Also it may influence such useful procedures as tensor reduction or Integration-by-Parts reduction of loop integrals, and can make them less straightforward to apply. In the context of SCET and TMD factorization, a few alternative approaches, preserving explicit Lorentz-covariance of the formalism, has been proposed. The analytic regularization is most convenient from the computational point of view and is widely applied in SCET calculations, but it has to be introduced on diagram-by-diagram basis and is used mostly in the context of real corrections~\cite{Becher:2011dz}. The proposal of $\delta$-regularization in a form of Ref.~\cite{Echevarria:2015byo} is also rather attractive and it is applicable to virtual corrections, but this approach modifies the standard definition of Wilson line in a nontrivial way (see Eq. 5 of Ref.~\cite{Echevarria:2015byo}), destroying their classical gauge-transformation properties, so in a moment it is not clear, if such modification is applicable in context of High-Energy EFT. Finally one should comment on the proposal of Ref.~\cite{vanHameren:2017hxx} to treat Eikonal propagators as standard propagators of fictitious particle with addition of large light-like momentum. If, as proposed in Ref.~\cite{vanHameren:2017hxx}, one expands the well-known results for scalar one-loop integrals in rapidity regulator variable {\it after} the expansion in $\epsilon=(4-d)/2$, where $d$ is the dimension of space-time, then one generally obtains double logarithms of rapidity regulator already at one loop, while from Reggeization one expects only single logarithms. Therefore double logarithms should cancel. In present work we first perform the asymptotic expansion in rapidity regulator and keep the exact in $\epsilon$ results up to a point when all rapidity divergences, except single logarithm, cancel away. Our analysis presented in in Sec.~\ref{Sec:1-loop-vert} shows, that this cancellation happens independently of the order of expansion in rapidity regulator vs. $\epsilon$, which might suggest that approach of Ref.~\cite{vanHameren:2017hxx} is also correct.  However, in the present paper we have chosen probably less technically convenient, but still tractable approach of {\it tilted Wilson lines}~\cite{CollinsQCD, Collins:2011ca,  Hentschinski:2011tz, Chachamis:2012cc, Chachamis:2012gh, Chachamis:2013hma}, which has the advantage that it does not modify gauge-transformation properties of Wilson lines and if applied properly -- allows one to preserve gauge-invariance of effective action. 

  Rapidity divergences in real and virtual corrections will be regularized if one shifts the directions of Wilson lines in the interaction part of the Lagrangian of EFT (\ref{Eq:L-Rg}, \ref{Eq:L-Qq}) from the light-cone, substituting:
\begin{eqnarray}
n^\pm_{\mu} \to \tilde{n}^{\pm}_\mu=n^{\pm}_\mu+r\cdot n^{\mp}_\mu,\ \ k_\pm \to \tilde{k}_{\pm}=k_{\pm}+r\cdot k_{\mp},\label{Eq:r-reg-def}
\end{eqnarray}        
 and therefore assigning finite rapidities $\mp \log r/2$ to the Wilson lines, where $0<r\ll 1$ is the rapidity regularization parameter and $\tilde{n}_{\pm}^2=4r$, $\tilde{n}_+\tilde{n}_-=2(1+r^2)$. This shift does not affect gauge-invariance of Lagrangian of interaction of Reggeized quarks with QCD quarks and gluons (\ref{Eq:L-Qq}). However, the Lagrangian (\ref{Eq:L-Rg}) has to be further modified. Indeed, after regularization, action contains the following term:
 \begin{eqnarray*}
  S_{Rg}\supset \int d^2 {\bf x}_T \int\limits_{-\infty}^{+\infty} \frac{dx_- dx_+}{2} {\rm tr} \left[ R_+(x) \tilde{\partial}_-  \partial_\rho^2 \tilde{W}_{\tilde{x}_+} [ \tilde{A}_- ] \right] = \int d^2 {\bf x}_T \int\limits_{-\infty}^{+\infty} \frac{d\tilde{x}_- d\tilde{x}_+}{1-r^2} {\rm tr} \left[ R_+(x) \frac{\partial}{\partial \tilde{x}_+}  \partial_\rho^2 \tilde{W}_{\tilde{x}_+} [ \tilde{A}_- ] \right],
 \end{eqnarray*}
  where we have passed from integration over $x_+$ and $x_-$ to integration over $\tilde{x}_+$ and $\tilde{x}_-$ with the Jacobian $1/(1-r^2)$ and we denote explicitly, that Wilson line depends on $\tilde{x}_+$. Integrating by parts over $\tilde{x}_+$ one obtains:
 \begin{eqnarray*}
 S_{Rg}\supset \int d^2 {\bf x}_T \int\limits_{-\infty}^{+\infty} \left. \frac{d\tilde{x}_-}{1-r^2} {\rm tr} \left[ R_+(x) \partial_\rho^2 \tilde{W}_{\tilde{x}_+} [ \tilde{A}_- ] \right] \right\vert_{\tilde{x}_+=-\infty}^{\tilde{x}_+=+\infty}  - \int d^2 {\bf x}_T \int\limits_{-\infty}^{+\infty} \frac{d\tilde{x}_- d\tilde{x}_+}{1-r^2} {\rm tr} \left[ \left(\tilde{\partial}_-R_+(x)\right)   \partial_\rho^2 \tilde{W}_{\tilde{x}_+} [ \tilde{A}_- ] \right].
\end{eqnarray*}
  
When $\tilde{x}_+=+\infty$, Wilson line in the first term stretches from $-\infty$ to $+\infty$ and such Wilson line is invariant w.r.t. gauge transformations which become trivial at large distances from the origin. Invariance w.r.t. this subgroup of gauge transformations is enough for perturbation theory in covariant gauges, which we will use. The second term in last expression is equal to zero at $r=0$ due to kinematic constraint (\ref{Eq:kin-constr-R}). To nullify it at $r\neq 0$ we propose to modify the kinematic constraints (\ref{Eq:kin-constr-R}) and (\ref{Eq:kin-constr-Q}) as follows:
  \begin{eqnarray}
  \tilde{\partial}_+ R_- = \tilde{\partial}_- R_+ = 0, \label{Eq:r-kin-constr-R}\\
  \tilde{\partial}_+ Q_- = \tilde{\partial}_- Q_+ = 0. \label{Eq:r-kin-constr-Q}
  \end{eqnarray}

As it was noted above, modification of kinematic constraint for Reggeized quarks is not necessary for gauge invariance, but it turns out, that calculation of many scalar integrals is simpler in such ``regularized'' MRK, so we propose to use constraint (\ref{Eq:r-kin-constr-Q}) at least on the level of scalar integrals. 

  It is instructive to understand first how the proposed regularization work on the level of real corrections. Let's consider the regularized Lipatov's $R_+gR_-$ vertex~\cite{BFKL1}:
\begin{eqnarray}
 \tilde{\Delta}^{abc}_{+\mu-}(q_1,q_2)= \tilde{\Delta}^{+\rho}_{ad}(q_1) \frac{-i}{q_1^2} \tilde{\Delta}_{-\rho\mu}^{dbc}(q_2,q_1) + \tilde{\Delta}_{-\mu\rho}^{abd}(q_2) \frac{-i}{q_2^2} \tilde{\Delta}^{+\rho}_{dc}(q_1,q_2-q_1) \nonumber \\ + \tilde{\Delta}^{+\rho_1}_{ad_1}(q_1) \frac{-i}{q_1^2} \tilde{\Delta}^{-\rho_2}_{cd_2}(q_2) \frac{-i}{q_2^2} (g_s f^{d_1 b d_2}) \gamma_{\rho_1\mu\rho_2}(q_1,q_2-q_1,-q_2) \nonumber \\
= g_s f^{abc} \left[ -(\tilde{n}_+\tilde{n}_-) \left( (q_1+q_2)_\mu + q_1^2 \frac{\tilde{n}^-_\mu}{\tilde{q}_2^-} + q_2^2\frac{\tilde{n}^+_\mu}{\tilde{q}_1^+} \right)+2\left( \tilde{q}_1^+\tilde{n}^-_\mu + \tilde{q}_2^-\tilde{n}^+_\mu \right) \right],\label{Eq:LV-reg}
\end{eqnarray} 
where $\gamma_{\mu_1\mu_2\mu_3}(k_1,k_2,k_3)=g_{\mu_1\mu_2}(k_1-k_2)_{\mu_3}+g_{\mu_2\mu_3}(k_2-k_3)_{\mu_1}+g_{\mu_1\mu_3}(k_3-k_1)_{\mu_2}$, momentum $q_1$ is incoming and $q_2$ is outgoing, so that the gluon momentum is $k=q_1-q_2$, and we have used the regularized kinematic constraints $\tilde{q}_1^-=\tilde{q}_2^+=0$. For the on-shell gluon with a given $k^+$ and $k^-$, the regularized kinematic conditions are satisfied by: $q_1^+=(k^++rk^-)/(1-r^2)$, $q_1^-=-rq_1^+$, $q_2^-=-(k^-+rk^+)/(1-r^2)$ and $q_2^+=-rq_2^-$. It is easy to verify, that vertex (\ref{Eq:LV-reg}) satisfies the Slavnov-Taylor identity $(q_1-q_2)^\mu\tilde{\Delta}^{abc}_{+\mu -}(q_1,q_2)=0$ for any value of $r$, while it would not be the case if modified kinematic conditions where not used. The square of regularized vertex is:
\[
(-g^{\mu\nu})\tilde{\Delta}^{abc_1}_{+\mu -}(q_1,q_2)\tilde{\Delta}^{abc_2}_{+\nu -}(q_1,q_2)=16N_c \delta_{c_1c_2}\frac{{\bf q}_{T1}^2 {\bf q}_{T2}^2}{{\bf k}_T^2} \left[f(y,r)+O(r) \right],
\]  
where we have neglected $O(r)$-terms in the numerator and introduced a function $f(y,r)=(re^{-y}+e^y)^{-2}(re^{y}+e^{-y})^{-2}$ of $y=\log(k^+/k^-)/2$. Function $f(y,r)$ provides a smooth cutoff at $\mp\log r/2$ to otherwise divergent integral over rapidity of a gluon:
\[
\int\limits_{-\infty}^{+\infty} dy\ f(y,r)= -\log r-1+O(r),
\]
so that rapidity divergence manifests itself as term $\sim\log r$.

\section{Rapidity divergences at one loop}
\label{Sec:RDs-gen}

  In this section we will derive general conditions for the appearance of rapidity divergences in one-loop scalar integrals with $n+1\geq 2$ external lines, among which $n_R=1$ or $2$ could be Reggeon lines. Integrals from higher-order induced vertices can have more than $n_R$ Eikonal propagators, but such integrals can be reduced to linear combination of integrals with at most $n_R$ Eikonal propagators by partial fractioning, so we will not consider this case. 

\begin{figure}
\begin{center}
\includegraphics[width=0.22\textwidth]{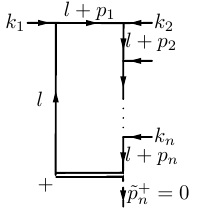}\hspace{2cm}
\includegraphics[width=0.28\textwidth]{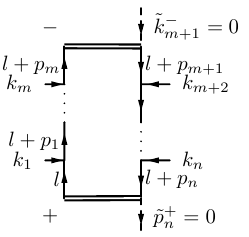}
\end{center}
\caption{Scalar one-loop integrals with one (left diagram) and two (right diagram) external Reggeon lines. Eikonal propagators are denoted by double lines. \label{Fig:1-loop}}
\end{figure}

 An example of integral with one Eikonal propagator $1/\tilde{l}^+$ is given in the left diagram in the Fig.~\ref{Fig:1-loop}, where momentum of the Reggeon satisfies modified kinematic constraint $\tilde{p}^+_n=0$. One notes, that it doesn't matter to which end of the Eikonal propagator one attaches the Reggeon line, since $\tilde{n}_+(l+p_n)=\tilde{l}^+$. Nontrivial one-loop integrals with two external Reggeon lines will contain at least two different Eikonal propagators, e.g. $1/\tilde{l}^+$ and $1/\tilde{l}^-$, and in this case, one can have $m\leq n-1$ external lines connected to the loop in between two Eikonal propagators, see the right diagram in Fig.~\ref{Fig:1-loop}. Momentum of the second Reggeon in this case satisfies the modified kinematic constraint $\tilde{k}_{m+1}^-=0$.  We denote the loop momentum as $l$ and momenta which flow through the propagators can be represented as $l+p_i$, $i=0,\ldots, n$, where $p_i=\sum\limits_{j=1}^i k_i,\ p_0=0$ and $k_i$, $i=1,\ldots,n$ are (incoming) momenta of external particles. In general, integrals which we are going to study can be written as:
\begin{eqnarray*}
I_1&=&\int\frac{d^d l}{(l^2)^{1-\alpha_0}((l+p_1)^2)^{1-\alpha_1}\ldots ((l+p_n)^2)^{1-\alpha_n} (\tilde{n}^+l+i\varepsilon)^{1-\beta_1}}, \\
I_2&=&\int\frac{d^d l}{(l^2)^{1-\alpha_0}((l+p_1)^2)^{1-\alpha_1}\ldots ((l+p_n)^2)^{1-\alpha_n} (\tilde{n}^+l+i\varepsilon)^{1-\beta_1}(\tilde{n}^-l+\tilde{p}^-_m+i\varepsilon)^{1-\beta_2}},
\end{eqnarray*}
where $d=4-2\epsilon$ and we have chosen the ``Euclidean'' pole prescriptions for the Eikonal poles, while the usual $+i\varepsilon$ prescription for quadratic propagators is implied. Integrals with other prescriptions for Eikonal poles can be related with this integrals by analytic continuation, as it will be discussed in Sec.~\ref{Sec:Ints}.

  For the general discussion of $I_1$ and $I_2$ it is convenient to employ the ``mixed'' version of Feynman's parametrization, where parameters for Eikonal propagators $x_{1,2}\in[0,+\infty)$, while parameters for the usual propagators $a_i\in[0,1]$, $i=0,\ldots, n$. Up to an overall factor one can write:
\begin{eqnarray*}
 I_1 \sim \int\limits_{0}^1 [d^{n+1}a] \int\limits_0^\infty \frac{dx_1}{x_1^{\beta_1}} \int d^dl\ \left[ \sum\limits_{i=0}^n a_i (l+p_i)^2 + x_1(\tilde{n}^+ l) \right]^{-n-2+\alpha+\beta_1}, \\
 I_2 \sim \int\limits_{0}^1 [d^{n+1}a] \int\limits_0^\infty \frac{dx_1}{x_1^{\beta_1}} \frac{dx_2}{x_2^{\beta_2}} \int d^dl\  \left[ \sum\limits_{i=0}^n a_i (l+p_i)^2 + x_1(\tilde{n}^+ l) + x_2(\tilde{n}^- l+\tilde{p}_m) \right]^{-n-3+\alpha+\beta},
\end{eqnarray*}
where $[d^{n+1}a]= \frac{da_0}{a_0^{\alpha_0}}\ldots \frac{da_n}{a_n^{\alpha_n}} \delta\left(1-\sum\limits_{i=0}^{n}a_i \right)$, $\alpha=\alpha_0+\ldots+\alpha_n$, $\beta=\beta_1+\beta_2$. After diagonalization of the quadratic form in square brackets and integration over $l$, $I_1$ and $I_2$ take the form:
\begin{eqnarray}
&&\hspace{-14mm}I_1\sim \int\limits_{0}^1 [d^{n+1}a] \int\limits_0^\infty \frac{dx_1}{x_1^{\beta_1}} \left[ {\cal D}+ x_1\sum\limits_{i=0}^{n} \tilde{p}^+_i a_i +rx_1^2 \right]^{-n-\epsilon-\alpha-\beta_1},\label{Eq:I1_par_r} \\
&&\hspace{-14mm}I_2\sim \int\limits_{0}^1 [d^{n+1}a] \int\limits_0^\infty \frac{dx_1}{x_1^{\beta_1}} \frac{dx_2}{x_2^{\beta_2}}\left[ {\cal D} + \sum\limits_{i=0}^{n}a_i(x_1\tilde{p}^+_i + x_2(\tilde{p}^-_i-\tilde{p}_m^-))+(1+r^2)x_1x_2 +r(x_1^2+x_2^2) \right]^{-n-1-\epsilon-\alpha-\beta}, \label{Eq:I2_par_r} 
\end{eqnarray}
where ${\cal D}=-\frac{1}{2}\sum\limits_{i,j=0}^{n} a_i a_j (p_i-p_j)^2$ is the usual parametric polynomial, corresponding to the same integral but without Eikonal propagators. 

  To study the rapidity divergences, one puts $r=0$, then integral over $x_1$ in $I_1$ and $x_2$ in $I_2$ can be easily calculated:
\begin{eqnarray}
&&\left. I_1^{(+)}\right\vert_{r=0}\sim \int\limits_{0}^1 [d^{n+1}a] \left(\sum\limits_{i=0}^{n} p^+_i a_i\right)^{-1+\beta_1} {\cal D}^{-n+1-\epsilon-\alpha-2\beta_1} , \label{Eq:I1_par_r-0}\\
&& \left. I_2^{(+-)}\right\vert_{r=0}\sim \int\limits_{0}^1 [d^{n+1}a] \int\limits_0^\infty \frac{dx_1}{x_1^{\beta_1}} \left(x_1+\sum\limits_{i=0}^{n}a_i(p^-_i-p_m^-) \right)^{-1+\beta_2} \left[{\cal D}+x_1 \sum\limits_{i=0}^{n-1}a_i p^+_i \right]^{-n-\epsilon-\alpha-\beta-\beta_2}  . \label{Eq:I2_par_r-0}
\end{eqnarray}

  In both cases, rapidity divergence arises from the factors in circular brackets. One considers first the case when powers of all propagators are equal to one, i.e. $\alpha_0=\ldots=\alpha_n=\beta_1=\beta_2=0$. Then for $I_1^{(+)}$ in the particular case $n=2$, due to kinematic condition $p^+_n=0$ the sum in circular brackets in Eq.~(\ref{Eq:I1_par_r-0}) consists only of one term: $p_1^+a_1$ and integral over $a_1$ is logarithmically divergent at $a_1\to 0$. This is rapidity divergence and it is not regularized by dimensional regularization, because ${\cal D}$ is finite when $a_1\to 0$. For $n>2$ no non-integrable singularity comes from the factor in circular brackets in Eq.~(\ref{Eq:I1_par_r-0}), because the corresponding sum will contain more than one term. Integrals with $n=0$ and $n=1$ are actually more singular and will be considered in Sec.~\ref{Sec:Ints-pow}.

  Judging from Eq.~(\ref{Eq:I1_par_r}) and (\ref{Eq:I1_par_r-0}), rapidity divergence for $n=2$ can be removed in three ways. First, one can keep $r$ finite. Then integral over $x_1$ doesn't produce a non-integrable singularity at $a_1\to 0$ as it will be shown in Sec.~\ref{Sec:Ints}. Second, one can keep $r=0$, but set $\beta_1>0$ or $\alpha_1<0$. This corresponds to analytic regularization of rapidity divergence. Third, one can differentiate $I_1$ w.r.t. $k_1^2$ or $k_2^2$, thus introducing additional factor $a_1$ into the numerator and so removing logarithmic divergence.  Therefore derivative of an integral w.r.t. external scale typically doesn't contain rapidity divergence and can be calculated at $r=0$. This convenient property will be exploited in Sec.~\ref{Sec:Ints-log}.   
   
  For the Eq.~(\ref{Eq:I2_par_r-0}) case $n=1$ is special. In this case $m=0$ and $p_1^-=0$ due to kinematic constraint of external Reggeon, so that the sum in circular brackets in Eq.~(\ref{Eq:I2_par_r-0}) consists only of one term: $x_1$, and integral over $x_1$ is logarithmically divergent for $x_1\to 0$ if powers of all propagators are equal to one. Since function ${\cal D}$ doesn't depend on $x_1$, this divergence can not be regularized by dimensional regularization. Again, divergence is absent if $r>0$ (``tilted Wilson line'' regularization) or if $r=0$ but $\beta_2>0$ or $\beta_1<0$ (analytic regularization). One notices, that  to analytically regularize the rapidity divergence, one always have to change powers of propagators forming the ``horizontal rungs'' of the ``ladder'', but this time both of this propagators are Eikonal. For $n>1$ there is no logarithmic divergence, because the sum in circular brackets in Eq.~(\ref{Eq:I2_par_r-0}) contains more than one term.  

  From the analysis above one concludes, that kinematic constraints (\ref{Eq:kin-constr-R}) and (\ref{Eq:kin-constr-Q}) are necessary for the appearance of rapidity divergences in loop integrals.

\section{Rapidity divergent scalar integrals}
\label{Sec:Ints}
\subsection{Scalar integrals with power rapidity divergences}
\label{Sec:Ints-pow}
  In this section we will consider scalar integrals with one external Reggeon line, and one or two quadratic propagators. We will use the following normalization of the measure of integration over loop momentum~\cite{1-loop-ints}:
 \[
[d^d l]=\frac{(\mu^2)^\epsilon d^d l}{i\pi^{d/2} r_\Gamma},\ r_\Gamma=\frac{\Gamma^2(1-\epsilon)\Gamma(1+\epsilon)}{\Gamma(1-2\epsilon)}=\frac{1}{\Gamma(1-\epsilon)}+O(\epsilon^3).
\]

  An integral with one Eikonal and one quadratic propagator is nonzero if there is an addition of external momentum $p$ in the quadratic propagator\footnote{Such integrals appear in process of tensor reduction}:
\begin{equation}
A_{[-]}(p)=\int\frac{[d^d l]}{(p+l)^2 [\tilde{l}^-]} = -\frac{\tilde{p}^-\ r^{-1+\epsilon} }{\cos(\pi \epsilon)} \frac{1}{2\epsilon (1-2\epsilon)} \left\lbrace \frac{\mu}{\tilde{p}^-} \right\rbrace^{2\epsilon}, \label{Eq:A0m}
\end{equation}
 where $\left\lbrace \frac{\mu}{k} \right\rbrace^{2\epsilon}=\frac{1}{2}\left[ \left(\frac{\mu}{k-i\varepsilon} \right)^{2\epsilon} + \left(\frac{\mu}{-k-i\varepsilon} \right)^{2\epsilon}   \right]$. Integral (\ref{Eq:A0m}) with $1/(\tilde{l}^-+i\varepsilon)$-propagator is Euclidean, and one obtains integral with PV pole prescription by analytic continuation in $\tilde{p}^-$. Integral (\ref{Eq:A0m}) contains rapidity divergence of the power type $\sim r^{-1+\epsilon}$. It is possible to treat such integrals in two ways. If one sets $r=0$ then integral (\ref{Eq:A0m}) is actually scaleless and one can put it to zero in dimensional regularization. But, as it will be shown below, if one keeps $r>0$, power divergences cancel between diagrams describing the same region in rapidity.

 An integral with two quadratic propagators:
\[
 B_{[-]}(p)=\int\frac{[d^d l]}{l^2 (p+l)^2 [\tilde{l}^-]},
\] 
will play an important role in our calculations

  For the non-light-like $p$ the leading terms of asymptotic expansion of this integral for $r\ll 1$ can be written as:
\begin{equation}
 B_{[-]}(p)= \frac{1}{p^- \epsilon^2} \left( \frac{\mu^2}{-p^2} \right)^\epsilon + \frac{1-2\epsilon }{\epsilon } \frac{rA_{[-]}(p)}{\tilde{p}_-^2} + \Delta B_{[-]}(-p^2, p_-) + O(r), \label{Eq:B0m}
\end{equation}
  where we have expressed the contribution $\sim r^{\epsilon}$ in terms of $A_{[-]}$ integral and $\Delta B_{[-]}$ is $O(r^{-\epsilon})$ contribution:
\begin{equation}
\Delta B_{[-]}(-p^2, p_-)=-\frac{1}{p_-}\left( \frac{p_-^2 \mu^2}{(-p^2)^2} \right)^\epsilon \frac{\Gamma^2(1-2\epsilon)\Gamma(1+2\epsilon)\cdot r^{-\epsilon}}{2\epsilon^2 \Gamma^2(1-\epsilon)}. \label{Eq:DB0m}
\end{equation}

  For light-like $p$ the first and third terms in Eq.~(\ref{Eq:B0m}) are absent. This terms are also absent for the special case when $p^2<0$ but $p^-=0$. Expression for $B_{[-]}$ published by us in Ref.~\cite{gaQq-real-photon} applies only in this special cases.  If $p^-=0$, asymptotic expansion of integral with Euclidean pole-prescription actually starts with the term $\sim r^{-1/2}$ (such term can be found e.g. in Appendix 1 of Ref.~\cite{Chachamis:2012cc}, integral [MSQ1]), but this term cancels in PV-prescription and asymptotic expansion of the integral with PV-prescription starts with $O(r^{\epsilon})$. Finally, in the $\tilde{p}^-=0$ case, the expansion of $B_{[-]}(p)$ actually starts with $O(r)$-term, and we put this integral to zero.

  The contribution (\ref{Eq:DB0m}) was obtained by resummation of infinite series of poles in the double Mellin-Barnes representation~\cite{Smirnov} for integral (\ref{Eq:B0m}). Due to the fact, that $1/\epsilon^2$-pole cancels, and parametric representation for $B_{[-]}$ is only a two-fold integral, the numerical cross-check of expressions (\ref{Eq:B0m}), (\ref{Eq:DB0m}) is rather straightforward and doesn't require specialized methods such as sector decomposition~\cite{Heinrich:2008si}. We have checked, that for sufficiently small $r$ the difference between numerical result for $B_{[-]}$ (with Euclidean pole-prescription) and analytic result (\ref{Eq:B0m}) scales as $O(r)$ independently on other parameters.

  Finally we will consider the ``tadpole'' and ``bubble'' integrals with two Eikonal propagators of the same type:
\[
A_{[--]}(p)=\int\frac{[d^d l]}{l^2 [\tilde{l}^-][\tilde{l}^- - \tilde{p}^-]},\  B_{[--]}(p)=\int\frac{[d^d l]}{l^2 (p+l)^2 [\tilde{l}^-] [\tilde{l}^-+\tilde{p}^-]},
\]
  which will arise in our calculation of $R{\cal O}g$ vertex. This integrals can be related to $A_{[-]}$ and $B_{[-]}$ by partial-fractioning of the Eikonal denominators. The results are:
\begin{equation}
A_{[--]}(p)=\frac{1}{\tilde{p}_-} A_{[-]}(p),\ B_{[--]}(p)=\frac{2}{\tilde{p}_-} B_{[-]}(p). \label{Eq:A0mm_B0mm}
\end{equation}

\subsection{Scalar integrals with logarithmic rapidity divergences}
\label{Sec:Ints-log}

As it follows from the analysis of Sec.~\ref{Sec:RDs-gen}, the simplest integral, containing logarithmic rapidity divergence is:
\[
B_{[+-]}(p)=\int\frac{[d^d l]}{l^2 (p+l)^2 [\tilde{l}^+] [\tilde{l}^-]},
\]
where $p_+=p_-=0$. Integral with PV-prescription for both poles can be expressed in terms of integral with Euclidean pole prescription:
\[
B_{(+-)}({\bf p}_T^2,r)=\int\frac{[d^d l]}{l^2 (p+l)^2 (\tilde{l}^++i\varepsilon) (\tilde{l}^-+i\varepsilon)},
\]  
as
\begin{equation}
B_{[+-]}(p)=\frac{1}{2}\left[B_{(+-)}({\bf p}_T^2,r) - e^{i\pi\epsilon} B_{(+-)}(-{\bf p}_T^2+i\varepsilon,-r+i\varepsilon)  \right].\label{Eq:Bpm_AC}
\end{equation}

  Parametric representation (with all Feynman parameters ranging from zero to infinity) for $B_{(+-)}$ has the form:
\[
B_{(+-)}({\bf p}_T^2,r)=\frac{\Gamma(2+\epsilon)(\mu^2)^\epsilon}{r_\Gamma} \int\limits_{0}^\infty dx_1 (1+x_1)^{2\epsilon} \int\limits_0^\infty dy_2 \int\limits_{r y_2}^{y_2/r} \frac{dy_3}{1-r^2}  \left[ x_1 {\bf p}_T^2 + y_2y_3 \right]^{-2-\epsilon}, 
\]
where we have made the change of variables $y_2=x_2+rx_3$, $y_3=x_3+rx_2$ with Jacobian $1/(1-r^2)$ in the original parametric representation. Taking the integral over $y_3$ one obtains, that $B_{(+-)}=(J(r)-J(1/r))/(1-r^2)$, with:
\[
J(r)=\frac{\Gamma(2+\epsilon)(\mu^2)^\epsilon}{r_\Gamma (1+\epsilon)} \int\limits_{0}^\infty dx_1 (1+x_1)^{2\epsilon} \int\limits_0^\infty \frac{dy_2}{y_2^{1-\alpha}} ({\bf p}_T^2 x_1 + ry_2^2)^{-1-\epsilon},
\]
where we have introduced intermediate regulator $\alpha>0$, dependence on which should cancel between $J(r)$ and $J(1/r)$. Intermediate regularization is needed since integral $B_{(+-)}$ converges only if the range of integration over $y_3$ is finite. But we wish to represent integral over finite range in $y_3$ as difference of two integrals $J(1/r)$ and $J(r)$ with $y_3$ extending to $+\infty$, so the intermediate regularization is necessary to make $J(r)$ well-defined. Integral $J(r)$ is easily calculated in terms of $\Gamma$-functions and the poles in $\alpha$ indeed cancel between $J(r)$ and $J(1/r)$ leading to the simple answer: $B_{(+-)}({\bf p}_T^2,r)=\left(\mu^2/{\bf p}_T^2 \right)^\epsilon 2\log r/{\bf p}_T^2/\epsilon$, which after substitution to the analytic continuation formula (\ref{Eq:Bpm_AC}) leads to
\begin{equation}
  B_{[+-]}({\bf p}_T^2)=\frac{1}{{\bf p}_T^2(1-r^2)}\left(\frac{\mu^2}{{\bf p}_T^2} \right)^\epsilon \frac{2\log r + i\pi}{\epsilon}. \label{Eq:B0pm}
  \end{equation} 

  Another integral with logarithmic rapidity divergence, which was identified in Sec.~\ref{Sec:RDs-gen}, is the integral with one external Reggeon line and three quadratic propagators:
\[
C_{[-]}(t_1, Q^2, q^-)=\int\frac{[d^d l]}{l^2 (q_1+l)^2 (q_1+q+l)^2 [\tilde{l}^-]},
\]
where $t_1=-q_1^2$, $Q^2=-q^2$, $(q+q_1)^2=0$ and $\tilde{q}_1^-=0$. We will first consider the case when $Q^2=0$, then on-shell condition for $q+q_1$ together with modified kinematic constraint for $q_1$ is satisfied by $q_1^+=t_1/{q_-}+t_1^2r/q_-^3+O(r^2)$, $q_1^-=-rq_1^+$ and parametric representation for $C_{[-]}$ with Euclidean pole prescription takes the form:
\[
C_{(-)}=-\frac{\Gamma(2+\epsilon)(\mu^2)^\epsilon}{r_\Gamma }  \int\limits_0^\infty dx_1 dx_2 dx_3\ (1+x_1+x_2)^{2\epsilon} \left[ (t_1+O(r))x_1 + x_3(x_2q_- + r x_3) \right]^{-2-\epsilon},
\]
  where $O(r)$-corrections do not depend on Feynman parameters and can be neglected, since they lead only to $O(r)$ corrections to the integral. Integral with PV-prescription is obtained from the latter integral as:
\begin{equation}
C_{[-]}(t_1,Q^2,q_-)=\frac{1}{2}\left[ C_{(-)}(t_1,Q^2,q_- -i\varepsilon) - C_{(-)}(t_1,Q^2,-q_- -i\varepsilon) \right]. \label{Eq:Cm_AC}
\end{equation}
 
 Anti-symmetrization of the integral $C_{(-)}$ w.r.t. light-cone component $q_-$ in Eq.~(\ref{Eq:Cm_AC}) can be understood as anti-symmetrization of the full amplitude w.r.t. the substitution $s\leftrightarrow -s$ and therefore the PV pole prescription indeed projects-out the part with definite signature from the amplitude. Integral $C_{(-)}$ can be straightforwardly expanded in $r\ll 1$ using one-fold Mellin-Barnes representation, which is not the case for parametric integral, obtained without modified kinematic constraint. Then the result with PV prescription is obtained using Eq.~(\ref{Eq:Cm_AC}): 
\begin{equation}
C_{[-]}(t_1,Q^2=0,q^-)=\frac{1}{q^- t_1} \left(\frac{\mu^2}{t_1} \right)^\epsilon \frac{1}{\epsilon} \left[ \log r+i\pi - \log\frac{|q_-|^2}{t_1} -\psi(1+\epsilon)-\psi(1)+2\psi(-\epsilon) \right] + O(r^{1/2}). \label{Ch2:eq:C0m_1-scale}
\end{equation}
 This integral was first calculated in Ref.~\cite{Hentschinski:2011tz}. 

  To find the $Q^2$-dependence of $C_{[-]}$ we will use the observation of Sec.~\ref{Sec:RDs-gen}, that derivative of it w.r.t. any scale will not contain rapidity divergence. Introducing parameter $X=Q^2/t_1$, taking the derivative of parametric representation w.r.t. $X$ and setting $r=0$ we obtain:
\[
\left.\frac{\partial C_{(-)}}{\partial X}\right\vert_{r=0}=\frac{t_1 \mu^{2\epsilon} \Gamma(3+\epsilon)}{r_\Gamma} \int\limits_0^\infty dx_1 dx_2 dx_3\  x_1 (1+x_1+x_2)^{2\epsilon}  \left[ t_1 x_1 (x_2+X) + q_- x_3 \right]^{-3-\epsilon}.
\]
  This integral can be straightforwardly taken by direct integration over $x_1$, $x_2$ and $x_3$. Introducing the function $I(X)= t_1 q_- \left(\frac{\mu^2}{t_1}\right)^{-\epsilon} \left[ C_{(-)}(X) - C_{(-)}(X=0) \right]_{r=0}$, with the boundary condition $I(0)=0$ one obtains:
\[
\frac{\partial I}{\partial X} = \frac{2 X^{-1-\epsilon}}{\epsilon} - \frac{2}{\epsilon} \frac{1-X^{-\epsilon}}{1-X}, 
\]
from where we find
\begin{eqnarray*}
I(X)&=&-\frac{2X^{-\epsilon}}{\epsilon^2} - \frac{2}{\epsilon} \int\limits_0^X \frac{(1-x^{-\epsilon}) dx}{1-x} = -\frac{2X^{-\epsilon}}{\epsilon^2} +2 \left[ -{\rm Li}_2(1-X)+\frac{\pi^2}{6} \right] + O(\epsilon).
\end{eqnarray*}

Leading $r$-dependence of the integral $C_{(-)}$ also contains the term $O(r^{-\epsilon})$. This term is invisible for the analysis above, because integral for $\partial C_{(-)}/\partial X$ converges at $\epsilon<0$ and this term vanishes for such $\epsilon$ in the limit $r\to 0$. As in the case of $B_{[-]}$-integral, this term can be calculated via resummation of a series of poles in the double Mellin-Barnes representation for the integral with $r\neq 0$. Up to a sign and dimensional factor $O(r^{-\epsilon})$ contribution to $C_{(-)}$ coincides with similar contribution to Eq.~(\ref{Eq:B0m}): $-\Delta B_{[-]}(Q^2,q_-)/t_1$, so that final result for $C_{[-]}$ at arbitrary values of $Q^2$ takes the form:
 \begin{equation}
C_{[-]}(t_1, Q^2, q_-)=C_{[-]}(t_1,Q^2=0,q_-) + \left(\frac{\mu^2}{t_1} \right)^\epsilon \frac{I(Q^2/t_1)}{q_- t_1} -\frac{1}{t_1}\Delta B_{[-]}(Q^2,q_-)+O(r^{1/2}). \label{Eq:C0m_2-scales}
\end{equation}

  Numerical cross-check of Eq.~(\ref{Eq:C0m_2-scales}) is more involved, because the parametric representation for this integral is three-fold and $\Delta B_{[-]}$ contribution introduces the $1/\epsilon^2$ pole. To numerically calculate an exact $r$-dependence of $C_{(-)}$ to all orders in $\epsilon$, we have implemented the sector-decomposition algorithm~\cite{Heinrich:2008si}. The integral in front of $1/\epsilon^2$ was calculated analytically and this contribution was subtracted from analytic result~(\ref{Eq:C0m_2-scales}) for comparison with numerical results. Numerically, with the help of the regular algorithm CUHRE of the CUBA library~\cite{CUBA}, we have calculated the exact $\epsilon$-dependence and $r$-dependencies only for $O(1/\epsilon)$ and finite parts of the integral $C_{(-)}$. It was found, that difference between analytic result (\ref{Eq:C0m_2-scales}) with $O(1/\epsilon^2)$ contribution subtracted and numerical results indeed scales as $O(r^{1/2})$ for sufficiently small $r$, which confirms Eq.~(\ref{Eq:C0m_2-scales}).    

\section{One-loop corrections to scattering vertices with two scales}
\label{Sec:1-loop-vert}
\subsection{One-loop correction to the $Q\gamma^\star q$-vertex}
\label{Sec:Q-ga-q-vert}

\begin{figure}
\begin{center}
\includegraphics[width=0.7\textwidth]{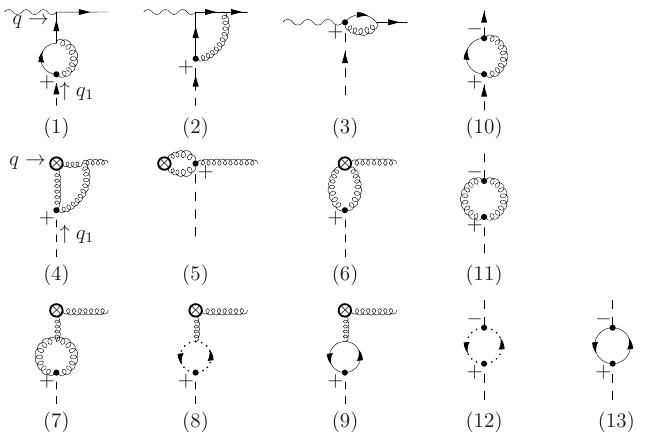}
\end{center}
\caption{Diagrams contributing to the one-loop correction to $Q\gamma^\star q$-vertex (1 -- 3), $R{\cal O}g$-vertex (4 -- 9), self-energies of Reggeized quark (10) and Reggeized gluon (11 -- 13). Dashed lines denote Reggeized gluons and quarks, dotted lines denote Faddeev-Popov ghosts. The vertex with a cross denotes an insertion of the operator ${\cal O}=-{\rm tr}[G_{\mu\nu}G^{\mu\nu}]/2$ with incoming momentum $q$.}\label{Fig:Q-ga-q_corr}
\end{figure}

  In this section we will calculate the one-loop correction $\Gamma^{(1)}_{+\mu}$ to the $Q_+\gamma^\star q$-vertex with the off-shell photon. It is given by EFT diagrams (1 -- 3) in the Fig.~\ref{Fig:Q-ga-q_corr}. We will denote incoming momentum of Reggeized quark as $q_1$, $q_1^2=-t_1$, $q_1^-=0$, incoming momentum of the photon as $q$: $q^2=-Q^2$ and outgoing quark is on-shell: $(q+q_1)^2=0$, so we will consider projection of this vertex on the on-shell spinor of QCD quark. To simplify the kinematics, we work in the reference frame, where off-shell photon have no transverse momentum, but has a ``large'' $q_-\gg \sqrt{Q^2}$ and ``small'' $q^+$ momentum components, so it's virtuality is given by $q^2=q_+q_-=-Q^2$.  

  The expressions for diagrams (1--3) are:
\begin{eqnarray}
{\cal M}_{\mu,ij}^{(1)}&=&e e_q \frac{g_s^2 C_F\delta_{ij}}{(2\pi)^D} \int d^d l\ \frac{\bar{u}(q+q_1)\gamma_\mu\hat{q}_1\gamma^{\rho} (\hat{q}_1+\hat{l})\tilde{\Gamma}^{(0)}_{+\rho}(q_1,l) \hat{P}_+}{q_1^2\ l^2\ (q_1+l)^2}, \label{Eq:M1} \\
{\cal M}^{(2)}_{\mu,ij}&=&e e_q \frac{g_s^2 C_F\delta_{ij}}{(2\pi)^D} \int d^d l\ \frac{\bar{u}(q+q_1)\gamma^\rho (\hat{q}_1+\hat{q}+\hat{l}) \gamma_\mu (\hat{q}_1+\hat{l})\tilde{\Gamma}^{(0)}_{+\rho}(q_1,l)\hat{P}_+ }{l^2\ (q_1+q+l)^2\ (q_1+l)^2}, \label{Eq:M2} \\
{\cal M}_{\mu,ij}^{(3)}&=&e e_q \frac{g_s^2 C_F\delta_{ij}}{(2\pi)^D} \int d^d l\ \frac{\bar{u}(q+q_1)\gamma^\rho (\hat{q}_1+\hat{q}+\hat{l}) \tilde{\Gamma}^{(0)}_{+\mu\rho}(q_1,q,l)\hat{P}_+ }{ l^2\ (q_1+q+l)^2}, \label{Eq:M3}
\end{eqnarray}
 where $i$ and $j$ are color indices of quarks, $C_F=(N_c^2-1)/2N_c$, $\tilde{\Gamma}^{(0)}_{+\rho}$ is the regularized $Q_+gq$-vertex (\ref{Eq:Qgq-vert}), $Q_+\gamma gq$-vertex $\tilde{\Gamma}^{(0)}_{+\mu\rho}(q_1,q,l)$ $=-\hat{q}_1(\tilde{n}_\mu^- \tilde{n}_\rho^-)/(q^- [\tilde{l}^-])$ can be derived from the $Q_+ggq$-vertex of the Lagrangian~(\ref{Eq:L-Qq}) and projector $\hat{P}_+$ (\ref{Eq:Proj-Qpm}) enforces constraint (\ref{Eq:kin_cons_Q2}) for the Dirac structure of $Q_+$-field. 

  For evaluation of diagrams (1 -- 3) we reduced tensor loop integrals down to scalar ones. Diagrams (1) and (2) can be separated into ``standard'' and ``Eikonal'' parts, where the latter contains Eikonal denominator $[\tilde{l}^-]$. The diagram (3) has only Eikonal part. Tensor reduction of standard parts can be performed by usual methods and tools, e.g. by routines of the FeynCalc package~\cite{FeynCalc}. Tensor reduction of integrals containing Eikonal denominator is completely analogous to this procedure, except, that now integral depends also on vector $\tilde{n}^-$, so this vector has to be added to the ansatz for the tensor structure. For example, for the rank-one three-point integral one has:
\begin{equation}
c_1\cdot (q_1+q)^{\rho} + c_2\cdot q_1^{\rho} + c_3\cdot \tilde{n}^{\rho}_-=\int \frac{d^d l \cdot l^{\rho}}{D_1 D_2 D_3 D_4}, \label{Eq:1-rank-ans}
\end{equation}  
where $D_1=(q_1+q+l)^2$, $D_2=(q_1+l)^2$, $D_3=l^2, D_4=\tilde{l}^-$. Contracting Eq.~(\ref{Eq:1-rank-ans}) with linearly-independent vectors which form the decomposition in the l.h.s. of this equation, one obtains system of linear equations for coefficients $c_i$. The r.h.s. of this system consists of linear combinations of integrals with scalar products in the numerator. All these scalar products can be expressed in terms of denominators $D_i$, thus leading to linear combinations of integrals with some denominators canceled. Hence, solving the linear system for $c_i$ one expresses them as linear combinations of scalar integrals with the same set denominators $\{D_i\}$ or it's subsets. At some stages of tensor reduction it is necessary to make shifts of the loop momentum by $q_1$ to restore $l^2$ denominator. We assume, that Eikonal denominator is insensitive to this shifts: $\tilde{l}^-\to \tilde{n}^-(l\pm q_1)=\tilde{l}^-$ due to modified kinematic constraint.  

  In case of the three-point function, the determinant of above-mentioned linear system (Gram determinant) is nonzero for $r\to 0$, so one can simplify the numerator of diagram (2), ignoring all $O(r)$ terms. This is also the case for diagram (3) but not for diagram (1) for which the Gram determinant is equal to $-4t_1r + O(r^2)$, so in this case one has to keep $O(r)$ terms in the numerator.

  The sum of three diagrams can be expressed as:
\[
{\cal M}_\mu^{(1)}+{\cal M}_\mu^{(2)}+{\cal M}_\mu^{(3)}= iee_q\bar{u}(q+q_1)\Gamma_{+\mu}^{(1)}(q_1,q),
\]  
  where one-loop correction to $Q\gamma^\star q$ vertex $\Gamma_{+\mu}^{(1)}(q_1,q)$ is naturally decomposed as:
\[
\Gamma_{+\mu}^{(1)}(q_1,q)=C[\Gamma]\cdot\Gamma_{+\mu}^{(0)}(q_1,q)+C[\Delta^{(1)}]\cdot\Delta_{+\mu}^{(1)}(q_1,q)+C[\Delta^{(2)}]\cdot\Delta_{+\mu}^{(2)}(q_1,q),
\] 
where $\Delta_{+\mu}^{(1)}(q_1,q)=\frac{\hat{q}}{q_-} \left( n_\mu^- - \frac{2(q_1)_\mu}{q_1^+} \right)$ and $\Delta^{(2)}_{+\mu}(q_1,q)=\frac{\hat{q}}{q_-} \left( n_\mu^- - \frac{q_\mu}{q^+} \right)$ . All three Dirac structures satisfy the QED Ward identities: $q^\mu \Gamma_{\mu}^{(0)}(q_1,q)=0$, $q^\mu \Delta_\mu^{(1)}(q_1,q)=0$, $q^\mu \Delta_\mu^{(2)}(q_1,q)=0$, but only $\Gamma_{+\mu}^{(0)}$ and $\Delta_{+\mu}^{(1)}$ where found for the case of real photon in Ref.~\cite{gaQq-real-photon}. The structure $\Delta_{+\mu}^{(2)}$ contributes only for non-zero $Q^2$ because coefficient in front of it decreases as $O(Q^2)$. Also, it doesn't contribute to the $F_2$ structure function of DIS.

  Coefficients in front of independent gauge-invariant structures are linear combinations of the following eight scalar integrals
\[
A_{[-]}(q),\ B(q),\ B(q_1),\ B_{[-]}(q),\ B_{[-]}(q_1),\ B_{[-]}(q+q_1),\ C(-q_1^2,-q^2),\ C_{[-]}(-q_1^2,-q^2,q^-),
\]
where we have introduced a short-hand notation for the standard one-loop bubble and triangle integrals of Ref.~\cite{1-loop-ints}: $B(q)=B(-q^2)=I_2^{D}(q^2;0,0)$, $C(-q_1^2,-q^2)=I^{D}_3(0,q_1^2,q^2;0,0,0)$. Expressions for the coefficients in terms of scalar integrals look as follows:
\begin{eqnarray}
 C[\Gamma]&=& -\frac{\bar{\alpha}_s C_F}{4\pi} \frac{1}{2} \left\{ \frac{[(d-8)Q^2+(d-6)t_1]B(t_1)-2(d-7)Q^2 B(Q^2)}{Q^2-t_1} \right. \nonumber \\
&& \left. -2\left[ (Q^2-t_1)C(t_1,Q^2)- q_- \left( t_1 C_{[-]}(t_1,Q^2,q_-)+(B_{[-]}(q)-B_{[-]}(q+q_1)) \right) \right] \right\}, \label{Eq:C-Gm} \\
 C[\Delta^{(1)}]&=& -\frac{\bar{\alpha}_s C_F}{4\pi} \frac{(Q^2+t_1) }{2(Q^2-t_1)^2}\left[ \left( (d-2)Q^2-(d-4)t_1 \right)B(t_1) -2Q^2 B(Q^2) \right] ,\label{Eq:C-D1} \\
 C[\Delta^{(2)}]&=& -\frac{\bar{\alpha}_s C_F}{4\pi} \frac{Q^2}{(Q^2-t_1)^2}\left[ \left( (d-6)t_1-(d-8)Q^2 \right)B(Q^2) +2(t_1-2Q^2) B(t_1) \right] ,\label{Eq:C-D2}
\end{eqnarray}  
were $\bar{\alpha}_s=\frac{\mu^{-2\epsilon} g_s^2}{(4\pi)^{1-\epsilon}} r_\Gamma$ is the dimensionless strong-coupling constant, and Gram-determinant singularity at $t_1=Q^2$ cancels if one substitutes expressions for scalar integrals.

 We observe the following pattern of cancellations of power rapidity divergences in the coefficients $C[\ldots]$. The coefficient in front of most singular integral $A_{[-]}(q)$ simplifies to zero if projector $\hat{P}_+$ in Eqns.~(\ref{Eq:M1}) -- (\ref{Eq:M3}) is in it's place, otherwise contribution of this integral is nonzero. The same cancellation happens with the coefficient in front of $B_{[-]}(q_1)$. Integrals $B_{[-]}(q)$ and $B_{[-]}(q+q_1)$ are contained only in $C[\Gamma]$ and coefficients in front of them {\it are equal and opposite in sign}, again if the projector $\hat{P}_+$ is present. This means, that $O(r^\epsilon)$ terms in Eq.~(\ref{Eq:B0m}) cancel between integrals $B_{[-]}(q+q_1)$ and $B_{[-]}(q)$, and only $Q^2$-dependent reminder of the latter integral is left. Apart from the $r$-independent term, above-mentioned reminder contains the term $\Delta B_{[-]}(Q^2,q_-)$ (\ref{Eq:DB0m}) which is $O(r^{-\epsilon})$. Since, the integral $B_{[-]}(q)$ stands together with $C_{[-]}(t_1,Q^2,q_-)$ in a combination $t_1 C_{[-]}(t_1,Q^2,q_-)+(B_{[-]}(q)-B_{[-]}(q+q_1))$ (see Eq.~(\ref{Eq:C-Gm})), the $\Delta B_{[-]}(Q^2,q_-)$-term cancels between this two integrals due to Eq.~(\ref{Eq:C0m_2-scales}). As a result, no power divergences or $O(r^{\pm\epsilon})$-terms is left and {the only rapidity divergence in one-loop correction to the $\Gamma^{(0)}_{+\mu}$ vertex is $O(\log r)$-divergence from integral $C_{[-]}$.} In particular, this fact means that $O(\log^2 r)$ terms which would show-up if we have expanded all integrals in $\epsilon$, should cancel-out. One should check such cancellation also in the approach of Ref.~\cite{vanHameren:2017hxx}. 

  Substituting explicit exact in $\epsilon$ results for Feynman integrals, one can take the limit $Q^2\to 0$ for $\epsilon<0$ and reproduce our results for on-shell photon case~\cite{gaQq-real-photon}, also our answer in this limit can be related with the results of Ref.~\cite{Kotsky:2002aq} for the one-loop correction to $Qgq$-vertex, obtained from QCD, however in this paper, gluon is on-shell, while quark is massive. After substitution of explicit expressions for scalar integrals and expansion in $\epsilon$, coefficients take the form:
 \begin{eqnarray}
 C[\Gamma]&=& -\frac{\bar{\alpha}_s C_F}{4\pi}\left\{\frac{1}{\epsilon^2} + \frac{L_1+1}{\epsilon} +  \left( \frac{1}{\epsilon} +\log \frac{\mu^2}{t_1} \right) (\log\bar{r}+i\pi) \right. \nonumber \\
&+&\left. \left[ -2{\rm Li}_2\left(1-\frac{Q^2}{t_1}\right) + L_1 + \frac{L_1^2-L_2^2}{2} - \frac{(2Q^2+t_1)L_2}{Q^2-t_1} - \frac{\pi^2}{6} + 3 \right]\right\}+O(r,\epsilon), \label{eq:C-Gm-Isub}\\
 C[\Delta^{(1)}]&=& -\frac{\bar{\alpha}_s C_F}{4\pi} \frac{(Q^2+t_1)(Q^2(L_2-1)+t_1)}{(Q^2-t_1)^2}+O(r,\epsilon), \label{eq:C-D1-Isub} \\
 C[\Delta^{(2)}]&=& -\frac{\bar{\alpha}_s C_F}{4\pi}\frac{2Q^2(Q^2-t_1-(2Q^2-t_1)L_2)}{(Q^2-t_1)^2}+O(r,\epsilon),
\end{eqnarray}
where $L_1=\log(\mu^2/Q^2)$, $L_2=\log(Q^2/t_1)$ and $\bar{r}= rQ^2/q_-^2$.

  The results obtained above can be used to compute one-loop correction to the partonic coefficient of $F_2$ DIS structure function in PRA. In this approach, one-loop correction to the partonic tensor, corresponding to subprocess $Q_+(q_1)+\gamma^\star(q)\to q$ has the form (see e.g. Eq. (5) in Ref.~\cite{NS_DIS1}):
\[
\hat{w}^{(1)}_{\mu\nu}=\frac{e^2e_q^2}{2}\cdot 2{\rm Re\ tr}\left[ (\hat{q}+\hat{q}_1) \Gamma^{(1)}_{+\mu} \frac{q_1^+ \hat{n}_-}{2} \Gamma^{(0)}_{+\nu}  \right],
\] 
where factor $1/2$ stands for the averaging over quark's spin. Projecting this tensor on the $F_2$ structure function, and dividing by the $d$-dimensional Born result for the partonic coefficient, one obtains the following expression for the one-loop correction factor to DIS partonic coefficient in PRA:
\begin{eqnarray}
C^{(1)}_{2q}(Q^2, t_1, \bar{r})=\frac{\bar{\alpha}_s C_F}{2\pi} \left\{-\frac{1}{\epsilon^2} -\frac{1+L_1}{\epsilon} -\left(\frac{1}{\epsilon} + \log\frac{\mu^2}{t_1} \right)\log\bar{r} +\left(\frac{\pi^2}{6}-2-L_1\right)\right. \nonumber\\
\left. +\frac{L_2^2-L_1^2}{2}+2{\rm Li}_2\left(1-\frac{Q^2}{t_1} \right) -\frac{1}{(Q^2-t_1)^2} \left[ Q^2(Q^2-t_1)+(t_1^2-2Q^4)L_2 \right] +O(r,\epsilon) \right\}. \label{eq:F2-CV}
\end{eqnarray}
  This result will be used for the calculation of $F_2$ structure function at NLO in PRA. This calculation will allow us to arrange the scheme of NLO calculations in PRA in such a way, that for single-scale observables, such as $F_2$, PRA will reproduce the NLO results in Collinear Parton Model up to higher-order in $\alpha_s(Q^2)$ and higher-twist terms, see Refs.~\cite{NS_DIS1, NS_DIS2} for more details.

\subsection{One-loop correction to $R {\cal O} g$-vertex}
\label{Sec:ORg-vert}

  In this section we will compute the one-loop correction to the $2\to 1$ scattering vertex induced by Reggeized gluon, with an insertion of gauge-invariant local operator:
\[
{\cal O}(x)=-\frac{1}{2}{\rm tr}\left[G_{\mu\nu}(x)G^{\mu\nu}(x) \right].
\] 
In momentum representation, the two-gluon vertex generated by operator ${\cal O}(x)$ reads:
\[
i\delta_{a_1 a_2}\left( (k_1k_2) g^{\mu_1\mu_2} - k_1^{\mu_2} k_2^{\mu_1}  \right),
\]
where momenta of gluons $k_{1,2}$ are incoming.

If this insertion is taken with a space-like momentum $q$ ($k_1+k_2+q=0$) while one final-state gluon is on-shell, kinematics of such process is identical to the partonic kinematics of DIS, and this process can be viewed an analog of DIS, directly sensitive to gluon PDF already in the LO in $\alpha_s$. This property is convenient for certain kinds of theoretical studies in perturbative QCD, see e.g. Refs.~\cite{Moch:H-DIS,Daleo:H-DIS} and references therein. We are planning to use this process to define the scheme of NLO calculations in PRA in the gluon channel, analogously to the case of quark-induced DIS subprocess, considered in the previous section. 

  Diagrams (4 -- 9) in the Fig.~\ref{Fig:Q-ga-q_corr} contribute to the one-loop correction which we are going to calculate. In this figure we have introduced for brevity the combined $R(q_1)g(k_1)g(k_2)$ and $R(q_1)g(k_1)g(k_2)g(k_3)$ vertices:
  \begin{eqnarray}
   V_{+\mu_1\mu_2}^{ab_1b_2}(k_1,k_2)&=&\tilde{\Delta}_{+\mu_1\mu_2}^{ab_1b_2}(q_1,k_1)  + \tilde{\Delta}^{+\rho}_{ad}(q_1) \frac{-i}{q_1^2} (g_s f^{d b_1 b_2}) \gamma_{\rho\mu_1\mu_2}(q_1,k_1,k_2), \label{Eq:RGG-vertex}\\
  && \tilde{\Delta}_{+\mu_1\mu_2\mu_3}^{ab_1b_2b_3}(q_1,k_1,k_2)  + \tilde{\Delta}^{+\rho}_{ad}(q_1) \frac{-i}{q_1^2} (-i g_s^2 ) \gamma^{db_1b_2b_3}_{\rho\mu_1\mu_2\mu_3}, \nonumber   
  \end{eqnarray}
where $\gamma^{db_1b_2b_3}_{\rho\mu_1\mu_2\mu_3}$ is the usual four-gluon vertex of QCD and induced vertex ${\Delta}_{+\mu_1\mu_2\mu_3}^{ab_1b_2b_3}$ is discussed in the \hyperlink{Sec:Appendix:A}{Appendix} and can be found, e.g. in Refs.~\cite{MH_PolePrescr, AntonovFRs}. The result for one-loop correction at hands is insensitive to details of the pole-prescription for this induced vertex, so we take all Eikonal poles in the PV-prescription for simplicity. Also we have introduced combined $Rq\bar{q}$ and $R$-ghost-$\overline{\rm ghost}$ vertices:
   \[
   \tilde{\Delta}^{+\rho}_{ad}(q_1) \frac{-i}{q_1^2} (ig_sT^d \gamma_\rho),\ \ \tilde{\Delta}_{+\rho}^{ad}(q_1) \frac{-i}{q_1^2} (-g_sf_{dbc} k_1^\rho),
   \]
so that expressions for diagrams (4 -- 9, 11 -- 13) in the Fig.~\ref{Fig:Q-ga-q_corr} can be straightforwardly written-down using vertices described above.  We work in the same reference frame as in Sec.~\ref{Sec:Q-ga-q-vert}, but now we have to satisfy the modified kinematic constraint (\ref{Eq:r-kin-constr-R}): $\tilde{q}_1^-=0$ due to a presence of Reggeized gluon. The solution of the on-shell condition for final-state gluon $(q+q_1)^2=0$ taken together with this constraint can be expanded in $r$ as follows:
\[
q_1^+=\frac{Q^2+t_1}{q_-}+\frac{r t_1 (Q^2+t_1)}{q_-^3} + O(r^2),
\]
and due to cancellations with $O(r)$ Gram-determinants, obtained amplitude will satisfy the Slavnov-Taylor identities of QCD up to one order in $r$ less, than number of orders in $r$ taken in the expansion for $q_1^+$.

  To reduce tensor integrals down to scalar ones, we apply the same strategy as in Sec.~\ref{Sec:Q-ga-q-vert}, but now, tensor integrals up to rank three arise, which are reduced sequentially to rank-two, rank-one and finally -- scalar integrals. In this process, the singular Gram-determinants $\sim O(r)$ can be found in the Eikonal part of the calculation, so one has to keep all terms of $O(r)$ in the numerator. In the standard part, $O(r)$-terms can be dropped from the beginning. 

  In the case of one-loop correction to $R{\cal O}g$-vertex we observe the similar pattern of cancellation of power-like rapidity divergences as in the case of $Q\gamma^\star q$-vertex. The contribution of $A_{[-]}(q)$ cancels between diagrams (4) and (5) due to the relation (\ref{Eq:A0mm_B0mm}). The $\sim r^{\epsilon}$ terms cancel between integrals $B_{[-]}(q)$ and $B_{[-]}(q+q_1)$ while the term $\sim r^{-\epsilon}$ cancels between $C_{[-]}(t_1,Q^2,q^-)$ and $B_{[-]}(q)$ all coming from diagrams (4) and (5), and relation (\ref{Eq:A0mm_B0mm}) is again important for this cancellations. As a result, only single-logarithmic rapidity divergence is left and the one-loop correction is proportional to the Born $R{\cal O}g$ vertex:
\begin{equation}
G^{(0)}_{+\mu}=\frac{i}{2} \left( (Q^2-t_1) n^-_\mu - 2q_- (q_1)_\mu \right).
\end{equation}

  In terms of scalar integrals, the $O(\alpha_s)$ coefficient reads:
\begin{eqnarray*}
C\left[ G_+^{(0)} \right] =\frac{\bar{\alpha}_s}{4\pi}\frac{1}{2} \left\{\frac{{B}({t_1})}{(d-2) (d-1)
   ({Q^2}-{t_1})^2} \left[{C_A} \left((d-2)
   (5 d-4) {Q}^4-2 (d (7 d-24)+16) {Q^2} {t_1} \right. \right. \right. \\
\left.\left. +(d-2) (5 d-4)
   {t_1}^2\right)-2 (d-2)^2 {n_F}
   ({Q^2}-{t_1})^2\right] -\frac{2 {C_A} (d-4) {Q^2} {B}({Q^2})}{(d-2)
   ({Q^2}-{t_1})^2} \left[
   (d-4) {Q^2}-(d-2) {t_1} \right] \\
\left. -2 {C_A} \left[{q_-} \left({t_1}
   {C_{[-]}}({t_1},{Q^2},{q_-})+{B_{[-]}}(q)-{B_{[-]}}(q+q_1)\right)+({t_1}-{Q^2})
   {C}({t_1},{Q^2})\right]\right\},
\end{eqnarray*}    
and after the substitution of scalar integrals and expansion in $\epsilon$, one has:
\begin{eqnarray*}
C\left[ G^{(0)} \right] =\frac{\bar{\alpha}_s}{4\pi} \left\{ -\frac{C_A}{\epsilon^2} + \frac{1}{\epsilon} \left[ \beta_0-C_A(1+L_1) \right]-C_A\left(\frac{1}{\epsilon} + \log\frac{\mu^2}{t_1} \right)  (\log\bar{r}+i\pi) \right. \\
+C_A\left[  2 {\rm Li}_2\left(1-\frac{Q^2}{t_1} \right) +\frac{L_2^2}{2} - L_2 - \frac{1}{2} L_1(L_1+2) + \frac{\pi^2}{6}-\frac{2}{3}  \right]
\left. +\beta_0 \left[ \frac{10}{6}+L_1+L_2 \right] + O(r,\epsilon) \right\},
\end{eqnarray*}
where $\beta_0=11C_A/3-2n_F/3$.

\subsection{One-Reggeon exchange. Comparison with QCD}
\label{Sec:comp-QCD}

  To check the one-loop corrections to $Q\gamma^\star q$ and $R{\cal O}q$ effective vertices, obtained above and demonstrate the self-consistency of EFT~\cite{Lipatov95, LV} we will compare Multi-Regge limit of specific one-loop QCD amplitudes with results obtained from EFT. To work with the scalar quantities, we will consider the interference terms between one-loop and tree-level amplitudes of the following subprocesses:
\begin{eqnarray}
\gamma^\star(q)+\gamma(P)\to q(q+q_1)+\bar{q}(P-q_1),\label{Eq:ga_ga-q_q} \\
{\cal O}(q)+g(P) \to g(q+q_1) + g(P-q_1), \label{Eq:O_g-g_g}
\end{eqnarray}
in massless QCD.  The photon indices in the interference of one-loop and tree-level amplitudes of the process (\ref{Eq:ga_ga-q_q}) will be projected on the $F_2$ structure-function. By Eq.~(\ref{Eq:O_g-g_g}) we understand the usual QCD amplitude $g\to g+g$ with an insertion of the local, gauge-invariant operator ${\cal O}$ carrying momentum $q$.

\begin{figure}
\begin{center}
\includegraphics[width=0.45\textwidth]{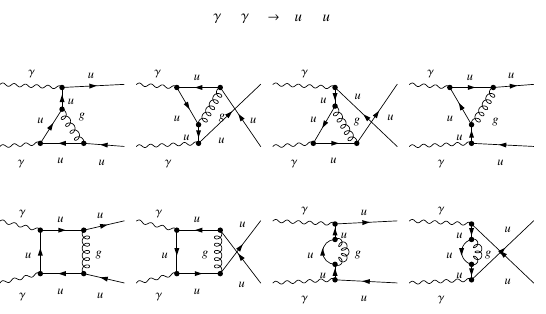} \\
\includegraphics[width=0.45\textwidth]{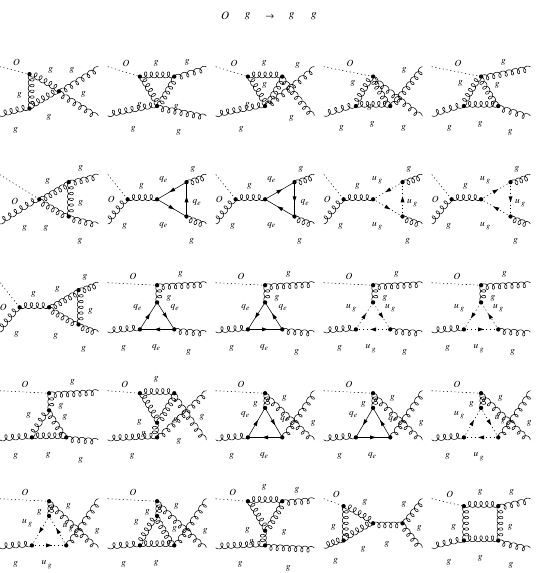} \includegraphics[width=0.45\textwidth]{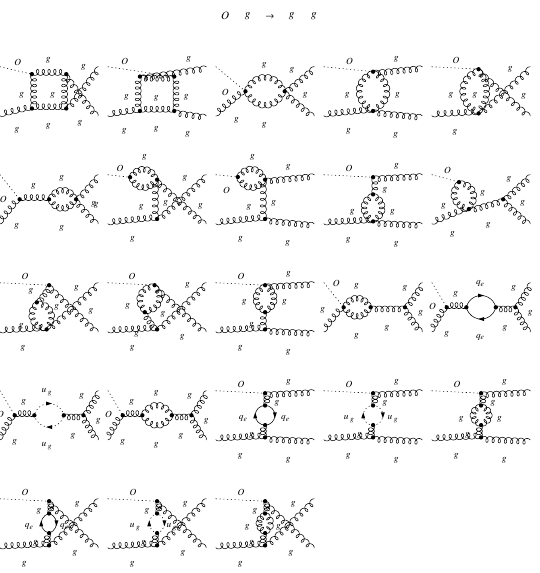}
\end{center}
\caption{Feynman diagrams contributing to one-loop correction to the amplitude of the process (\ref{Eq:ga_ga-q_q}) (top panel) and (\ref{Eq:O_g-g_g}) (bottom panel) in massless QCD in dimensional regularization. Dotted line denotes the insertion of the operator ${\cal O}=-{\rm tr}[G_{\mu\nu}G^{\mu\nu}]/2$ with incoming momentum $q$. \label{Fig:1-loop-QCD}}
\end{figure}

  For convenience, we will work in the center-of-mass frame of momenta $P$ and $q$. In terms of usual Bjorken variable $x_B=Q^2/2(Pq)$, light-cone components of $P$ and $q$ can be expressed in this frame as:
\begin{eqnarray*}
q^+=-x_B P_+,\ q^-=\frac{Q^2}{x_B P_+},\ {\bf q}_T=0;\ 
P_+=\sqrt{\frac{Q^2}{x_B (1-x_B)}},\ P_-={\bf P}_T=0,
\end{eqnarray*}
so that in the Regge limit $x_B\to 0$ : $P^+\to \infty$, $q^-\to\infty$, while $q^+\to 0$.  We have calculated the interferences of one-loop amplitudes of the processes (\ref{Eq:ga_ga-q_q}) and (\ref{Eq:O_g-g_g}), given by diagrams in the Fig.~\ref{Fig:1-loop-QCD}, with their tree-level counterparts. We use FeynArts~\cite{FeynArts} to generate Feynman diagrams, FeynCalc~\cite{FeynCalc} to perform index contractions, color algebra and express all scalar-products as linear combinations of denominators, then remaining scalar integrals with positive and negative powers of denominators are reduced by LiteRed~\cite{LiteRed1, LiteRed2} to the small set of master integrals. For the process (\ref{Eq:ga_ga-q_q}) we just contract the indices of real photons, while indices of virtual photon we project on the $F_2$ structure-function, using the projector:
\[
P_2^{\mu\nu}=\frac{Q^2}{(d-2) (Pq)} \left[ -g^{\mu\nu} + Q^2 (d-1)\frac{P^\mu P^\nu}{(Pq)^2}  \right].
\]

 For the process (\ref{Eq:O_g-g_g}) we sum over polarizations of initial and final-state gluons in Feynman gauge and then add appropriate diagrams where all gluon cut-loops are replaced by ghost cut-loops. Out of such diagrams, only those with gluon in $t$-channel on both sides of a cut contribute to the Regge limit. 

 We divide-out the $d$-dimensional Born results and expand obtained expressions in the limit $x_B\to 0$ (after expansion in $\epsilon$), the leading term of this expansion for the process (\ref{Eq:ga_ga-q_q}) is:
\begin{eqnarray}
F^{\rm (1,QCD)}_{2}(Q^2, t_1, x_B)=\frac{\bar{\alpha}_sC_F}{4\pi}\left\{  -\frac{2}{\epsilon^2} -\frac{3}{\epsilon}-\frac{2L_1}{\epsilon} + \left( \frac{2\pi^2}{3}-7-L_1^2-3L_1 \right) +\right. \nonumber \\
+ \left(\frac{1}{\epsilon}+\log\frac{\mu^2}{t_1}\right)\left(\log \frac{1}{x_B^2}-2\pi i \right) +L_2^2+2{\rm Li}_2\left(1-\frac{Q^2}{t_1}\right) \nonumber  \\
\left. - \frac{1}{(Q^2-t_1)^2}\left[ Q^2(Q^2-t_1)+(3t_1^2-4Q^2 t_1)L_2 \right] \right\}+O(\epsilon, x_B), \label{eq:C_2-QCD}
\end{eqnarray}
while for the process (\ref{Eq:O_g-g_g}) it is:
\begin{eqnarray}
G^{\rm (1,QCD)}(Q^2, t_1, x_B)=\frac{\bar{\alpha}_sC_A}{4\pi}\left\{ -\frac{3}{\epsilon^2} -\frac{1}{\epsilon}(3L_1+L_2) + \left(\frac{1}{\epsilon} + \log\frac{\mu^2}{t_1} \right) \left(\log\frac{1}{x_B^2}-i\pi \right) \right.\nonumber\\
\left. +2{\rm Li_2}\left(1-\frac{Q^2}{t_1} \right) + \frac{1}{2}(L_2-3L_1)(L_1+L_2) +\frac{2\pi^2}{3}  \right\}+O(\epsilon, x_B), \label{eq:G-QCD}
\end{eqnarray}
and this QCD results should be predicted by EFT.

\begin{figure}
\begin{center}
\includegraphics[width=0.42\textwidth]{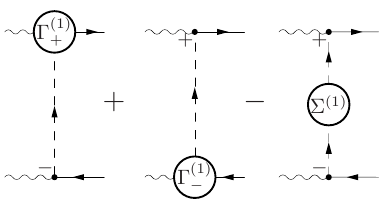}\hfill
\includegraphics[width=0.42\textwidth]{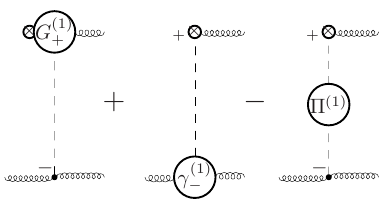}
\end{center}
\caption{One-Reggeon exchange contributions to the Regge limits of amplitudes of the process (\ref{Eq:ga_ga-q_q}) (left panel) and (\ref{Eq:O_g-g_g}) (right panel).}\label{Fig:EFT-1-R}
\end{figure}

  In this section we will consider the EFT contributions with one Reggeon in $t$-channel to amplitudes computed above (Fig.~\ref{Fig:EFT-1-R}). They can be constructed from the one-loop corrections to $Q\gamma^\star g$ and $R{\cal O}g$-vertices, computed in Sec.~\ref{Sec:Q-ga-q-vert} and Sec.~\ref{Sec:ORg-vert}, one-loop corrections to $Q\gamma q$ and $Rgg$-vertives, computed respectively in Refs.~\cite{gaQq-real-photon} and~\cite{Chachamis:2012cc, Nefedov:2019mws} and one-loop corrections to propagators of Reggeized gluon and quark, which where computed respectively in Refs.~\cite{Chachamis:2012cc} and~\cite{gaQq-real-photon}. The propagator corrections are given by diagrams (10) and (11--13) in the Fig.~\ref{Fig:Q-ga-q_corr} and can be expressed in terms of one-loop ``bubble'' and integral~(\ref{Eq:B0pm}):
\begin{eqnarray}
\Sigma^{(1)}(q_1)&=&\frac{\bar{\alpha}_s C_F}{4\pi} \frac{i\hat{q}_1}{2}\left[ (d-6) B(t_1)-2t_1 B_{[+-]}(q_1) \right], \\
\Pi^{(1)}(q_1)&=& \frac{\bar{\alpha}_s}{4\pi} \frac{t_1}{d-1} \left[ \left( (2-3d)C_A+2(d-2)n_F \right)B(t_1)+2C_A(d-1)t_1B_{[+-]}(q_1) \right].
\end{eqnarray}

  The minus sign in front of the Reggeon self-energy contributions in the Fig.~\ref{Fig:EFT-1-R} appears due to the necessity to ``localize'' one-loop corrections to effective vertices in rapidity~\cite{Hentschinski:2011tz,Chachamis:2012cc,Chachamis:2012gh,Chachamis:2013hma,gaQq-real-photon}. As it was noted in the end of Sec.~\ref{Sec:EFT}, our regularization for rapidity divergences in real corrections acts as a smooth cutoff at rapidities $\pm\log(1/r)/2$. It is instructive to extend this point of view also to the case of virtual corrections. Then the following picture emerges. Rapidity of the loop momentum in the ``backward''-vertex (e.g. $\Gamma_{+\mu}^{(1)}$) is naturally bounded from below by some scale $\sim -\log q_-/\sqrt{t_1}$, while, due to regularization of $1/l^-$-pole, it is also bounded from above by $\log(1/r)/2$. For the ``forward'' vertex situation is reversed: rapidity of the gluon in the loop is bounded from above by the natural rapidity scale of this vertex $\sim\log P_+/\sqrt{t_1}$ while from below it is cut-off by $-\log(1/r)/2$ due to regularization of $1/l^+$ pole. Finally for the ``central'' contribution, which is given by Reggeon self-energy, containing both poles, rapidity of momentum in the loop is bounded from below by $-\log (1/r)/2$ and from above by $\log(1/r)/2$. To make logarithmic divergences cancel we should remove double counting of the region $-\log (1/r)/2\leq y \leq \log (1/r)/2$ from contributions of ``forward'' and ``backward'' vertices, i.e. we should subtract the central contribution from each of them. In total, we are subtracting the central contribution two times and then adding it back, thus we get a minus sign in front of this contribution in the Fig.~\ref{Fig:EFT-1-R}. As a result of this procedure, logarithmic rapidity divergences cancel between ``forward'', ``backward'' and ``central'' contributions and we are left with $r$-independent result for the Regge limit. 

  To compare the EFT results with QCD results (\ref{eq:C_2-QCD}) and (\ref{eq:G-QCD}) we contract our Regge limit of one-loop amplitudes, obtained from EFT with the corresponding Born results and perform the projection on the $F_2$-structure function in the photon case. We have found, that EFT reproduces both real and imaginary parts of Eq.~(\ref{eq:G-QCD}) exactly, while for Eq.~(\ref{eq:C_2-QCD}) only the real and {\it one half} of imaginary part comes out right from our one-Reggeon exchange EFT calculation\footnote{The same is true for imaginary part of amplitude studied in our Ref.~\cite{gaQq-real-photon}. In Eq.~(21) of this reference, the imaginary part of one-Reggeon exchange amplitude is written and our statement that it accounts for the full imaginary part of the amplitude is wrong. Instead one should consider also two Reggeon exchange diagrams as we do in Sec.~\ref{Sec:Imag}.}.

 The reason for this mismatch is that one-Reggeon exchange diagrams, which we have taken into account, are responsible only for the {\it positive-signature part} of the amplitude of the process (\ref{Eq:ga_ga-q_q}), while the part with negative signature is given by diagrams with two Reggeons in $t$-channel, as it will be shown in the next section.

\subsection{Two-Reggeon exchange. Imaginary part}
\label{Sec:Imag}

\begin{figure}
\includegraphics[width=0.725\textwidth]{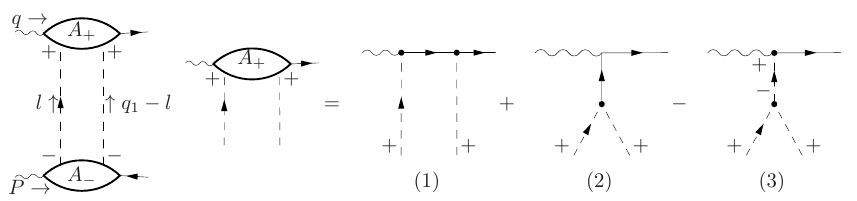} \\
\includegraphics[width=\textwidth]{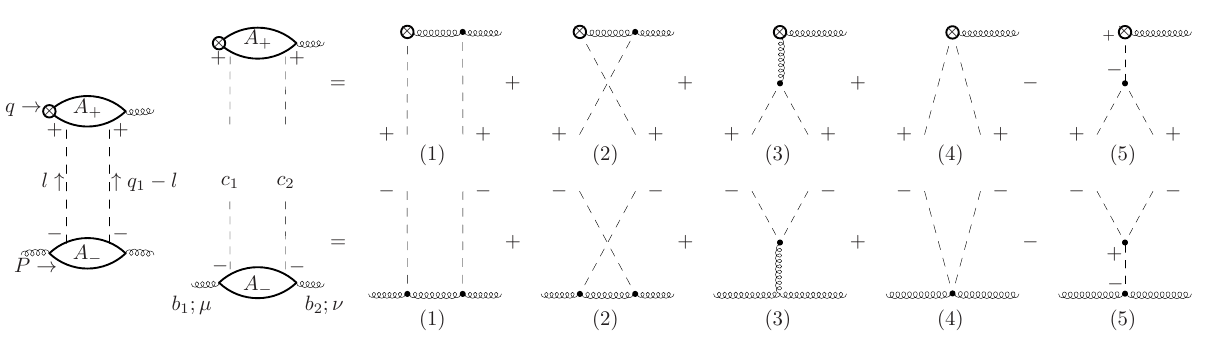}
\caption{Factorization of two-Reggeon contribution into a product of impact-factors, connected by Reggeon propagators and the structure of impact-factors for the subprocesses (\ref{Eq:ga_ga-q_q}) (top panel) and (\ref{Eq:O_g-g_g}) (bottom panel). For the quark case, the impact-factor $A_-$ contains diagrams of the same topology as in $A_+$, up to $+\leftrightarrow -$ exchange.}\label{Fig:2-R}
\end{figure}

  In the EFT~\cite{Lipatov95,LV}, negative-signature part of the amplitude of the process (\ref{Eq:ga_ga-q_q}) at the perturbative order we are considering is given by two-Reggeon contribution, depicted diagrammatically in the top panel of Fig.~\ref{Fig:2-R}. This quantity naturally factorizes into a product of two impact-factors, connected by propagators of Reggeized quark and gluon:
\begin{equation}
{\cal M}_{\mu\rho}^{(QR)}=\frac{g_s^2 C_F}{2} (ee_q)^2 \int  \frac{d^{d-2}{\bf l}_T}{(2\pi)^d}\frac{A_\mu^+(l_T)\hat{l}_T A_\rho^-(l_T)}{{\bf l}_T^2 ({\bf q}_{T1} - {\bf l}_T)^2} ,\label{Eq:M_QR-fact}
\end{equation}
where an overall factor $1/2$ comes from the propagator of Reggeized gluon~(\ref{Eq:L_kin}). In Eq.~(\ref{Eq:M_QR-fact}) we have decomposed loop-momentum integration measure as $d^{d}l=d^{d-2}{\bf l}_T (dl_+ dl_-)/2$ and moved integrations over light-cone components $l_+$ and $l_-$ into the corresponding impact-factors. 

 The impact-factor $A_+$ is given by three diagrams, shown in the Fig.~\ref{Fig:2-R}. Contribution of the diagram (1) is:
\[
A_\mu^{(+,1)}=i\int\limits_{-\infty}^{+\infty}\frac{dl_+}{\sqrt{2}}\frac{\bar{u}(q+q_1) \hat{n}_- \left(q_-\hat{n}_+/2+ (l_++q_+)\hat{n}_-/2+\hat{l}_T\right)\Gamma^{(0)}_{+\mu}(l,q) \hat{P}_+}{q_-l_+ -{\bf l}_T^2+i\varepsilon},
\] 
  where the first factor $\hat{n}_-$ in the numerator has come from the $R_+q\bar{q}$-vertex: $iT^b \gamma^\rho\cdot (-i/q^2)\cdot\Delta^{ab}_{+\rho}(q)=-iT^a \hat{n}_-$ and due to the MRK kinematic constraint, the $l^-$-component of the loop momentum does not propagate into $A_+$ impact-factor so denominator of the quark propagator does not depend on it. The numerator of $A_+$ simplifies into 
\[
2q_- \bar{u}(q+q_1) \hat{P}_+ \gamma_\mu \hat{P}_+ = 2q_- \bar{u}(q+q_1) \Gamma_{+\mu}^{(0)}(q_1,q) \hat{P}_+,
\]
  so it doesn't depend neither on $l^+$ nor on ${\bf l}_T$ and for the contribution of the first diagram we obtain the representation:
\begin{equation}
A_\mu^{(+,1)}= \left( \bar{u}(q+q_1) \Gamma_{+\mu}^{(0)}(q_1,q) \hat{P}_+\right)\int\limits_{-\infty}^{+\infty} \frac{i\sqrt{2}\ dl_+}{l_+-{\bf l}_T^2/q_- + i\varepsilon },\label{Eq:Ap_1}
\end{equation} 
which is ill-defined by itself and become convergent only when combined with other diagrams in the Fig.~\ref{Fig:2-R}. 

  Contribution of the diagram (2) in the Fig.~\ref{Fig:2-R} is equal to zero, because the $Q_+ q R_+$-vertex:
\[
\Delta_{+\rho}(q_1-l)\frac{-ig^{\rho\sigma}}{(q_1-l)^2}\Gamma_{+\sigma}^{(0)}(l,q_1-l)\hat{P}^+=0.
\] 

  Finally we have to ``localize in rapidity'' the impact-factor $A_+$ to avoid double-counting with the Regge-pole (positive-signature) contribution. To this end we subtract diagram (3) in the Fig.~\ref{Fig:2-R} from the impact-factor. Similar procedure applies e.g. in case of $qR_+R_+q$ impact-factor (see Sec. 3.2 of Ref.~\cite{MHThesis}) where analogous subtraction removes the double counting between the contribution with two Reggeized-gluons in $t$-channel (positive signature) and one Reggeized gluon contribution (negative signature). The diagram (3) gives:
\begin{eqnarray}
A_\mu^{(+,3)}&=& -i \int\limits_{-\infty}^{+\infty}\frac{dl_+}{\sqrt{2}}\bar{u}(q+q_1) \Gamma_{+\mu}^{(0)}(q_1,q)\hat{P}_+ \frac{\hat{q}_1}{q_1^2} \hat{P}_+\Gamma^{(0)}_{++-}(l,l-q_1,-q_1)\hat{P}_+ \nonumber \\
&=& \left( \bar{u}(q+q_1) \Gamma_{+\mu}^{(0)}(q_1,q) \hat{P}_+\right) \int\limits_{-\infty}^{+\infty} \frac{i\sqrt{2}\ dl_+}{[l^+-q_1^+]}, \label{Eq:Ap_3}
\end{eqnarray}
where the $Q_+R_+Q_-$ vertex $\hat{P}_+ \Gamma^{(0)}_{++-}(q_1,q_2,q_3)\hat{P}_+$ is obtained from $Q_+gQ_-$-vertex (\ref{Eq:FS-vertex}) as $\hat{P}_+ \Delta_{+\rho}(q_2) \times \frac{-ig^{\rho\sigma}}{q_2^2}\times $ $\Gamma_{+\sigma-}^{(0)}(q_1,q_2,q_3)\hat{P}_+ $ $= 2\hat{q}_{T3}/[l_+]$. Combining Eq.~(\ref{Eq:Ap_1}) and (\ref{Eq:Ap_3}) together one obtains:
\begin{eqnarray*}
A_\mu^{(+)}&=&A_{\mu}^{(+,1)}-A_{\mu}^{(+,3)}= \left( \bar{u}(q+q_1) \Gamma_{+\mu}^{(0)}(q_1,q) \hat{P}_+\right) \int\limits_{-\infty}^{+\infty} \frac{i\sqrt{2}\ dl_+ \left[ (l_+-q_1^+)\left( -q_1^++{\bf l}_T^2/q_-+i\varepsilon \right)+O(\varepsilon^2) \right]}{\left(l_+-{\bf l}_T^2/q_- + i\varepsilon \right) (l_+-q_1^++i\varepsilon) (l_+-q_1^+-i\varepsilon)} \\
&=&  \left( \bar{u}(q+q_1) \Gamma_{+\mu}^{(0)}(q_1,q) \hat{P}_+\right)\times \pi\sqrt{2},
\end{eqnarray*}
which is independent on ${\bf l}_T$. Analogously for the second impact-factor one obtains
\[
A_\rho^{(-)}=\left( \hat{P}_+  \Gamma_{-\rho}^{(0)}(-q_1,P) v(P-q_1) \right)\times(-\pi\sqrt{2}),
\]
so for the two-Reggeon contribution we have:
\begin{eqnarray}
{\cal M}_{\mu\rho}^{(QR)}&=& {\cal M}_{\mu\rho}^{(0)}  \frac{g_s^2 C_F}{2} \int \frac{d^{d-2}{\bf l}_T}{(2\pi)^d} \frac{\hat{l}_T\hat{q}_{T1}  (-2i\pi^2)}{{\bf l}_T^2 ({\bf q}_{T1} - {\bf l}_T)^2} ={\cal M}_{\mu\rho}^{(0)}\times (-i\pi)\frac{\bar{\alpha}_s C_F}{4\pi} \frac{1}{\epsilon}\left(\frac{\mu^2}{{\bf q}_{T1}^2} \right)^\epsilon\nonumber \\
&=&{\cal M}_{\mu\rho}^{(0)}\times(-i\pi)\frac{\bar{\alpha}_s C_F}{4\pi} \left(\frac{1}{\epsilon}+\log\frac{\mu^2}{t_1}+O(\epsilon) \right),\label{Eq:neg-sign-Im}
\end{eqnarray}
which gives exactly the half of imaginary part of expression (\ref{eq:C_2-QCD}), missing in the single-Reggeon contribution to the amplitude of subprocess (\ref{Eq:ga_ga-q_q}), which was computed in the Sec.~\ref{Sec:comp-QCD}.

  Now we switch to the two-Reggeon exchange contribution to the amplitude of the subprocess (\ref{Eq:O_g-g_g}). This contribution also factorizes into a product of impact-factors, which is shown diagrammatically in the bottom panel of Fig.~\ref{Fig:2-R}:
\begin{equation}
{\cal M}_{\mu\nu\sigma}^{(RR),b_1b_2b_3}=\frac{1}{8} \int \frac{d^{d-2}{\bf l}_T}{(2\pi)^d} \frac{ A_\sigma^{+,c_2c_1b_3}(l_T) A_{\mu\nu}^{-,b_1c_1c_2b_2}(l_T) }{{\bf l}_T^2 ({\bf q}_{T1}-{\bf l}_T)^2},\label{Eq:M-RR}
\end{equation}
where $b_1, \mu$ are color and Lorentz indices of incoming gluon, $b_2,\nu$ and $b_3,\sigma$ correspond to outgoing gluons and into the factor $1/8$ in Eq.~(\ref{Eq:M-RR}) we have collected two factors of $1/2$ from Reggeon propagators and an additional factor $1/2$ which compensates for the double-coting of diagrams, arising because we have rewritten the sum of diagrams as a product of two impact-factors. 
   
  Let's consider the most interesting case of the $gR_-R_-g$ impact-factor $A^-$. Contribution of the diagram (1) contains two $Rgg$-vertices (\ref{Eq:RGG-vertex}) and can be written as: 
\begin{equation}
A_{\mu\nu}^{(-,1),b_1c_1c_2b_2}= (-i g_s^2) \int\limits_{-\infty}^{+\infty} \frac{dl_-}{\sqrt{2}} \frac{f^{b_1c_1d} f^{dc_2b_2}}{(P-l)^2}\times V_{-\mu\rho}(P,l-P) V_{-\rho\nu}(P-l,q_1-P). \label{Eq:gRRg-IF-0}
\end{equation}

  The $Rgg$-vertex (\ref{Eq:RGG-vertex}) can be decomposed as:
\begin{eqnarray}
V_{-\mu\nu}(k_1,k_2)&=&v_{-\mu\nu}(k_1,k_2)+ \left( k_{2\nu} n^+_\mu - k_{1\mu}n^+_\nu  +\frac{k_1^2 + k_2^2}{k_1^+} n^+_\mu n^+_\nu  \right), \label{Eq:RGG-dec}\\
v_{-\mu\nu}(k_1,k_2)&=&2\left[-k_1^+ g_{\mu\nu} - k_{2\mu} n_\nu^+ + k_{1\nu}n_\mu^+ + \frac{k_1k_2}{k_1^+}n_\mu^+ n_\nu^+  \right],\label{Eq:RGG-v-def}
\end{eqnarray}
where the vertex (\ref{Eq:RGG-v-def}), used e.g. in Ref.~\cite{Hentschinski:2018rrf}, has several convenient properties:
\begin{eqnarray}
&& n_\mu^+ v_{-\mu\nu}(k_1,k_2)=0,\ n_{\nu}^+ v_{-\mu\nu}(k_1,k_2)=0, \label{Eq:prop:1} \\
&& k_1^\mu v_{-\mu\nu}(k_1,k_2)=0,\ k_2^\nu v_{-\mu\nu}(k_1,k_2)=0, \label{Eq:prop:2} \\
&& v_{-\mu\rho}(k_1,k_2)v_{-\rho\nu}(-k_2,k_3)=-2k_1^+ v_{-\mu\nu}(k_1, k_3), \label{Eq:prop:3}
\end{eqnarray}
which hold for any values of $k_1^2$ and $k_2^2$. Using properties (\ref{Eq:prop:1}), (\ref{Eq:prop:2}) and omitting the terms proportional to $P_\mu$ and $(P-q_1)_\nu$, because they will be nullified after the contraction with polarization vectors of external gluons, one can put the product of two $Rgg$-vertices in the Eq.~(\ref{Eq:gRRg-IF-0}) into a following form:
\begin{equation}
V_{-\mu\rho}(P,l-P) V_{-\rho\nu}(P-l,q_1-P) = v_{-\mu\rho}(P,l-P) v_{-\rho\nu}(P-l,q_1-P)- (P-l)^2 n_\mu^+ n_\nu^+ . \label{Eq:VV-dec}  
\end{equation}
The first term in Eq.~(\ref{Eq:VV-dec}) can be reduced to $v_{-\mu\nu}(P,P-q_1)$ by the property (\ref{Eq:prop:3}), while the second ``anomalous'' term contains the factor $(P-l)^2$, which will cancel the gluon propagator.

 The similar ``anomalous'' term, with the color structure $f^{b_1c_2d}f^{dc_1b_2}$ comes from the diagram (2). The diagram (3) is equal to zero for the same reason as diagram (2) in the quark case, and diagram (4) cancels ``anomalous'' terms from diagrams (1) and (2), because it is given just by the $n^+$-projection of the QCD four-gluon vertex and is equal to:
\begin{equation}
A_{\mu\nu}^{(-,4),b_1c_1c_2b_2}= (-ig_s^2)  \int\limits_{-\infty}^{+\infty} \frac{dl_-}{\sqrt{2}} \left( f^{b_1c_1d}f^{dc_2b_2}+f^{b_1c_2d}f^{dc_1b_2} \right) n^+_\mu n^+_\nu .
\end{equation}

Therefore the sum of diagrams (1), (2) and (4) is equal to:
\begin{eqnarray}
A_{\mu\nu}^{(-,1+2+4),b_1c_1c_2b_2}&=&(-ig_s^2) v_{-\mu\nu}(P,P-q_1) \int\limits_{-\infty}^{+\infty} \frac{dl_-}{\sqrt{2}} \times \nonumber\\
&& \left\{ \frac{ f^{b_1\{c_1d}f^{dc_2 \}b_2} }{2} \left[\frac{(-2P_+)}{-P_+ l_- -{\bf l}_T^2 + i\varepsilon } + \frac{(-2P_+)}{P_+ l_- +2{\bf l}_T{\bf q}_{T1}-{\bf l}_T^2 + i\varepsilon } \right] \right. \nonumber\\
&+& \left. \frac{f^{b_1[ c_1d}f^{dc_2] b_2}}{2} \left[ \frac{(-2P_+)}{-P_+ l_- -{\bf l}_T^2 + i\varepsilon } - \frac{(-2P_+)}{P_+ l_- +2{\bf l}_T{\bf q}_{T1}-{\bf l}_T^2 + i\varepsilon } \right] \right\}, \label{Eq:Am-before-subtr}  
\end{eqnarray}
where we have introduced symmetric and anti-symmetric color structures: $f^{b_1\{c_1d}f^{dc_2 \}b_2}=f^{b_1c_1d}f^{dc_2b_2}+f^{b_1c_2d}f^{dc_1b_2}$, $f^{b_1[c_1d}f^{dc_2] b_2}=f^{b_1c_1d}f^{dc_2b_2}-f^{b_1c_2d}f^{dc_1b_2}=f^{b_1 d b_2} f^{d c_1 c_2}$ by Jacobi identity. Integral in front of the symmetric color-structure is finite, because $l_-$-term in the numerator cancels (see also Sec. 3.2 of Ref.~\cite{MHThesis}), but this color-structure in the $A^-$ impact-factor doesn't contribute to the amplitude of the process (\ref{Eq:O_g-g_g}), because the color-structure of impact-factor $A^+$ in this case is anti-symmetric.   

  The integral in front of anti-symmetric color-structure in Eq.~(\ref{Eq:Am-before-subtr}) is ill-defined, because as in the case of subprocess (\ref{Eq:ga_ga-q_q}) we have to perform a localization of impact-factor in rapidity by subtracting from it the diagram (5) with internal Reggeized-gluon line. This diagram contains the $R_-R_+R_-$-vertex: $V_{+--}^{abc}(q_1,q_2,q_3)=g_s f^{abc} q_1^2(n_+n_-)^2/[q_2^-]$, so after the subtraction integral over $l^-$ in color anti-symmetric part of Eq.~(\ref{Eq:Am-before-subtr}) becomes:
\[
\int\limits_{-\infty}^{+\infty}\frac{dl_-}{\sqrt{2}} \left\{ \frac{1}{l_- + {\bf l}_T^2/P_+ - i\varepsilon} + \frac{1}{l_-+(2{\bf l}_T {\bf q}_{T1} - {\bf l}_T^2)/P_+ + i\varepsilon } - \frac{1}{l_--i\varepsilon} - \frac{1}{l_-+i\varepsilon} \right\} = 0.
\]

  Hence in the case of the process (\ref{Eq:O_g-g_g}), the two-Reggeon exchange amplitude doesn't contribute at one loop and in agreement with this, the QCD result (\ref{eq:G-QCD}) is reproduced completely by one-Reggeon exchange contribution, discussed in Sec.~\ref{Sec:comp-QCD}.

\section{Conclusions}
\label{Sec:Conclusions}

  In the present paper the one-loop corrections to $Q\gamma^\star q$ and $R{\cal O}g$ effective vertices has been computed (Sec.~\ref{Sec:1-loop-vert}) using the formalism of gauge-invariant EFT for Multi-Regge processes in QCD~\cite{Lipatov95, LV} with ``tilted Wilson lines'' prescription for regularization of rapidity divergences. For this sake, scalar integrals $C_{[-]}$ and $B_{[-]}$ with additional scales of virtuality has been computed (Sec.~\ref{Sec:Ints}) for the first time. Consistency of the obtained results with QCD has been verified by comparison of Regge limits of $\gamma^\star +\gamma\to q+\bar{q}$ and ${\cal O}(q)+g\to g+g$ amplitudes computed in the EFT and directly in QCD in Sec.~\ref{Sec:comp-QCD} and two-Reggeon contributions has been studied in the Sec.~\ref{Sec:Imag}. These results will be instrumental in development of Parton Reggeization Approach~\cite{NS_PRA,NS_DIS1,NS_DIS2} or other techniques of NLO calculations in $k_T$-factorization~\cite{vanHameren:2017hxx} and serve as a nontrivial cross-check of the formalism of effective action for Reggeized gluons~\cite{Lipatov95} and quarks~\cite{LV} for the case of amplitudes containing more than one scale of virtuality.  

\section*{Acknowledgements}

  Author would like to acknowledge Joachim Bartels, Andreas van Hameren, Martin Hentschinski, Bernd Kniehl and Vladimir Saleev for multiple discussions of different aspects of loop calculations in the framework of Lipatov's EFT and Alexander von Humboldt foundation for awarding him with the Research Fellowship for Postdoctoral Researchers. This work also has been supported in part by RFBR grant \# 18-32-00060.

\section*{Appendix : Higher-order induced vertices from the Hermitian Reggeon-gluon interaction}\hypertarget{Sec:Appendix:A}{}

  In the Ref.~\cite{MH_PolePrescr} the pole-prescription for higher-order induced vertices has been proposed, satisfying the following properties: the $R_{\pm}g\ldots g$ induced vertex should be Bose-symmetric w.r.t. QCD gluons and should not depend on a sign of $\varepsilon$ in the $i\varepsilon$ prescription. The latter property ensures that the single-Reggeon exchange in the EFT indeed possesses a definite signature. In this appendix we will show, that effective vertices, which can be derived from the Hermitian $Rg$-Lagrangian~(\ref{Eq:L-Rg}) satisfy both properties. In a moment we do not have the general proof of this statement, so we check it for $O(g_s^2)$ and $O(g_s^3)$ induced vertices.

  Above we have shown, that Lagrangian~(\ref{Eq:L-Rg}) automatically leads to the PV-prescription for the Eikonal pole in the the $R_+gg$ vertex.  The $R_+ggg$ and $R_+gggg$ induced vertices, generated by the Lagrangian~(\ref{Eq:L-Rg}) look as follows:
\begin{eqnarray}
\Delta_{+\mu_1\mu_2\mu_3}^{ab_1b_2b_3}=-ig_s^2q^2 (n_{\mu_1}^- n_{\mu_2}^- n_{\mu_3}^-) \sum\limits_{(i_1,i_2,i_3)\in S_3} \frac{{\rm tr}\left[T^a \left(T^{b_{i_1}}T^{b_{i_2}} T^{b_{i_3}} +  T^{b_{i_3}}T^{b_{i_2}} T^{b_{i_1}}\right) \right]}{(k_{i_3}^- +i\varepsilon)(k_{i_3}^-+k_{i_2}^-+i\varepsilon)}, \label{Eq:R+_g3-vert} \\
\Delta_{+\mu_1\mu_2\mu_3\mu_4}^{ab_1b_2b_3b_4}=-ig_s^3q^2 (n_{\mu_1}^- n_{\mu_2}^- n_{\mu_3}^- n_{\mu_4}^-) \sum\limits_{(i_1,i_2,i_3,i_4)\in S_4} \frac{{\rm tr}\left[T^a \left(T^{b_{i_1}}T^{b_{i_2}} T^{b_{i_3}} T^{b_{i_4}} - T^{b_{i_4}} T^{b_{i_3}}T^{b_{i_2}} T^{b_{i_1}}\right) \right]}{(k_{i_4}^- +i\varepsilon)(k_{i_4}^-+k_{i_3}^-+i\varepsilon) (k_{i_4}^-+k_{i_3}^-+k_{i_2}^-+i\varepsilon)}, \label{Eq:R+_g4-vert}
\end{eqnarray} 
where $q$ is the incoming momentum of Reggeon, $k_{1,\ldots,4}$ are incoming momenta of QCD gluons, the summation goes over permutations of three or four objects and due to the kinematic constraint (\ref{Eq:kin-constr-R}) the $(-)$-component of momentum of gluons is conserved: $k_1^-+k_2^-+k_3^-=0$ and $k_1^-+k_2^-+k_3^-+k_4^-=0$ for $O(g_s^2)$ and $O(g_s^3)$ vertices respectively. 

  In Ref.~\cite{MH_PolePrescr} the $W^\dagger$-term in Eq.~(\ref{Eq:L-Rg}) is not included into the Lagrangian. Instead, the color structure of induced vertices obtained from the $Rg$-Lagrangian without $W^\dagger$-term is decomposed over the certain symmetry basis and it is shown, that only terms in the subspace spanned by ``maximally-nested'' commutators, which in the notation of Ref.~\cite{MH_PolePrescr} are defined as:
\[
[[[i_1,i_2],i_3],i_4]={\rm tr}\left\{T^a \left[\left[[T^{b_{i_1}},T^{b_{i_2}}], T^{b_{i_3}}\right], T^{b_{i_4}} \right] \right\} = \frac{-i}{2}f^{b_{i_1}b_{i_2}c_1} f^{c_1 b_{i_3},c_2} f^{c_2 b_{i_4} a},
\]
have desired properties. Decomposing the color structure of Eqns.~(\ref{Eq:R+_g3-vert}) and (\ref{Eq:R+_g4-vert}) over the same basis, one discovers, that only surviving color structures are the same nested commutators, and $S_3(1,2,3)$-structure for the case of Eq.~(\ref{Eq:R+_g3-vert}) or $S_3([i_1,i_2],i_3,i_4)$-structures for the Eq.~(\ref{Eq:R+_g4-vert}). Using the identities:
\begin{eqnarray*}
&& \frac{1}{k_{i_1}^-\pm i\varepsilon}\frac{1}{k_{i_1}^-+k_{i_2}^-\pm i\varepsilon} + \frac{1}{k_{i_2}^-\pm i\varepsilon}\frac{1}{k_{i_1}^-+k_{i_2}^-\pm i\varepsilon} = \frac{1}{k_{i_1}^-\pm i\varepsilon}\frac{1}{k_{i_2}^-\pm i\varepsilon},\\
&& \frac{1}{k^-+i\varepsilon}-\frac{1}{k^--i\varepsilon}=-2\pi i \delta(k^-),
\end{eqnarray*}
which hold in a distributional sense, one can  show, that the part of induced vertices (\ref{Eq:R+_g3-vert}) and (\ref{Eq:R+_g4-vert}) which is proportional to nested commutators reproduces results of Ref.~\cite{MH_PolePrescr}, while remnants, proportional to other color structures can be simplified as follows:
\begin{eqnarray}
\Delta_{+\mu_1\mu_2\mu_3}^{ab_1b_2b_3}&=&-ig_s^2q^2 (n_{\mu_1}^- n_{\mu_2}^- n_{\mu_3}^-)\left\{ [{\rm nested\ comm.}]-\frac{4}{3}\pi^2\delta(k_1^-)\delta(k_2^-) S_3(1,2,3) \right\}, \label{Eq:R+_g3-vert_simpl} \\
\Delta_{+\mu_1\mu_2\mu_3\mu_4}^{ab_1b_2b_3b_4}&=&-ig_s^3q^2 (n_{\mu_1}^- n_{\mu_2}^- n_{\mu_3}^- n_{\mu_4}^-)\left\{ [{\rm nested\ comm.}] \right. \nonumber\\
&-&8\pi^2\left[ d(k_2^-)\delta(k_3^-)\delta(k_4^-) S_3([1,2],3,4)  + d(k_3^-)\delta(k_2^-)\delta(k_4^-) S_3([1,3],2,4) \right. \nonumber\\
&&\hspace{5mm}  + d(k_4^-)\delta(k_2^-)\delta(k_3^-) S_3([1,4],2,3) + d(k_3^-)\delta(k_1^-)\delta(k_4^-) S_3([2,3],1,4) \nonumber\\ 
&&\hspace{4mm} \left. \left. +  d(k_4^-)\delta(k_1^-)\delta(k_3^-) S_3([2,4],1,3)+d(k_4^-)\delta(k_1^-)\delta(k_2^-) S_3([3,4],1,2) \right] \right\}, \label{Eq:R+_g4-vert_simpl}
\end{eqnarray}
where distribution $d(k)=1/(k-i\varepsilon)-i\pi\ {\rm sgn}(\varepsilon)\delta(k)$ has the property $d(-k)=-d(k)$, which ensures the Bose-symmetry of the induced vertex if one takes into account the conservation of $k^-$-momentum component. We have checked, that the color-symmetric terms in Eq.~(\ref{Eq:R+_g3-vert_simpl}) and (\ref{Eq:R+_g4-vert_simpl}) are independent on the sign of $\varepsilon$ and therefore are compatible with definite signature of one-Reggeon exchange.

  If one assumes the modified kinematic constraints (\ref{Eq:r-kin-constr-R}), the regularization of rapidity divergences in Eqns.~(\ref{Eq:R+_g3-vert_simpl}) and (\ref{Eq:R+_g4-vert_simpl}) boils down to replacements $k_i^-\to \tilde{k}_i^-$.

  In Ref.~\cite{Chachamis:2013hma} the rapidity-divergent part of two-loop correction to the Reggeized gluon propagator has been computed, using the pole-prescription of Ref.~\cite{MH_PolePrescr}, and consistency with known results for two-loop Regge trajectory of a gluon has been demonstrated. However, there is no contradiction between this result and Eq.~(\ref{Eq:R+_g3-vert_simpl}), since additional term in Eq.~(\ref{Eq:R+_g3-vert_simpl}) does not contribute to the rapidity-divergence of Reggeon propagator at two loops. In this term the Eikonal propagators are replaced by delta-functions, which kill the logarithmic rapidity divergence, which could have come from each loop of the two-loop diagram (${\rm k}_1$) in the Fig.~4 of Ref.~\cite{Chachamis:2013hma}. Therefore this term contributes neither to $O(\alpha_s^2 \log^2 r)$ nor to $O(\alpha_s^2 \log r)$ parts, but possibly contributes to the finite part of the two-loop correction to the Reggeon propagator, which was beyond the scope of Ref.~\cite{Chachamis:2013hma}. Therefore neither prescription of Ref.~\cite{MH_PolePrescr} nor usage of the Hermitian form of $Rg$-Lagrangian~(\ref{Eq:L-Rg}) is favored by existing calculations and further comparisons between scattering amplitudes in EFT and QCD are required to decide which option is the correct one.  


\bibliography{mybibfile}

\end{document}